\documentclass[manuscript,screen]{acmart}

\usepackage{multirow}       %
\usepackage{colortbl}       %
\usepackage[abbreviations]{glossaries-extra}    %
\usepackage{afterpage}      %
\usepackage{adjustbox}      %
\usepackage{tikz}           %
\usepackage{makecell}       %
\usepackage{caption}
\newcommand{\mytilde}{\raise.17ex\hbox{$\scriptstyle\mathtt{\sim}$}}
\newcommand{\rc}{\rowcolor{gray!10}}

\glsdisablehyper        %

\setabbreviationstyle{long-short}
\newabbreviation{se}{SE}{Software Engineering}
\newabbreviation{re}{RE}{Requirements Engineering}
\newabbreviation{its}{ITS}{Issue Tracking System}
\newabbreviation{po}{PO}{Product Owner}

\setabbreviationstyle[long]{long}
\newabbreviation[category=long]{ta}{}{Thematic Analysis}
\newabbreviation{mgr}{}{manager}
\newabbreviation{dev}{}{developer}

\setcopyright{cc}
\setcctype{by}
\acmJournal{TOSEM}
\acmYear{2026} \acmVolume{1} \acmNumber{1} \acmArticle{}
\acmMonth{1} \acmDOI{10.1145/3788876}

\thanks{© 2026 owner/author(s). This is the author's version of The Work. It is posted here for your personal use. Not for redistribution. The definitive version was published in ACM TOSEM Journal, \url{https://doi.org/10.1145/3788876.}}

\begin{document}

\title{Smells Depend on the Context: An Interview Study of Issue Tracking Problems and Smells in Practice}

\author{Lloyd Montgomery}
\email{lloyd.montgomery@uni-hamburg.de}
\orcid{0000-0002-8249-1418}
\author{Clara Lüders}
\email{clara.marie.lueders@uni-hamburg.de}
\orcid{0000-0001-7743-4067}
\author{Christian Rahe}
\email{christian.rahe@uni-hamburg.de}
\orcid{0009-0007-2110-6663}
\author{Walid Maalej}
\email{walid.maalej@uni-hamburg.de}
\orcid{0000-0002-6899-4393}
\affiliation{%
  \institution{University of Hamburg}
  \city{Hamburg}
  \country{Germany}
}

\begin{abstract}
\glspl{its} enable software developers and managers to collect and resolve issues collaboratively.
While researchers have extensively analysed \gls{its} data to automate or assist specific activities such as issue assignments, duplicate detection, or priority prediction, developer studies on \glspl{its} remain rare.
Particularly, little is known about the challenges \gls{se} teams encounter in \glspl{its} and when certain practices and workarounds (such as leaving issue fields like ``priority'' empty) are considered problematic.
To fill this gap, we conducted an in-depth interview study with 26 experienced \gls{se} practitioners from different organisations and industries.
We asked them about general problems encountered, as well as the relevance of 31 \gls{its} smells (aka potentially problematic practices) discussed in the literature.
By applying \gls{ta} to the interview notes, we identified 14 common problems including issue findability, zombie issues, workflow bloat, and lack of workflow enforcement.
Participants also stated that many of the \gls{its} smells do not occur or are not problematic.
Our results suggest that \gls{its} problems and smells are highly dependent on context factors such as \gls{its} configuration, workflow stage, and team size.
We also discuss potential tooling solutions to configure, monitor, and visualise \gls{its} smells to cope with these challenges.
\glsresetall  %
\end{abstract}

\ccsdesc[500]{Software and its engineering~Programming teams}
\ccsdesc[500]{Software and its engineering~Agile software development}
\ccsdesc[500]{Software and its engineering~Software maintenance tools}

\keywords{Issue Tracking Systems, Developer Experience, Developers Studies, Issue Trackers, Issue Linking}

\received{30 June 2025}
\received[revised]{10 November 2025}
\received[accepted]{6 January 2026}

\maketitle

\section{Introduction}

\glspl{its} are essential tools for software development projects.
Issues within \glspl{its} represent \gls{se} artefacts such as requirements~\cite{Montgomery:MSR:2022}, development tasks~\cite{VanCan_2024_REFSQ}, defect reports~\cite{Bettenburg:FSE:2008,Zimmermann:TSE:2010}, support tickets~\cite{Montgomery_2017_RE}, and improvement requests~\cite{Montgomery:MSR:2022}.
Much of development work today is organised along these issues~\cite{Bertram:CSCW:2010,Lucassen_2016_FSE}.
For instance, user stories get planned, implemented, and tested following workflows defined along issue states.
Defects get assigned, discussed, and fixed based on corresponding issue reports.
Often, development and maintenance tasks are also captured as issues to plan the work and keep track of the progress.
Modern \glspl{its} serve as project-, backlog-, and release-management tools.
Stakeholders exchange knowledge in comments sections and coordinate their work through notifications and workflow features~\cite{Arya:ICSE:2019}.
It is difficult to think of a modern software organisation without thinking of its Jira, RedMine, or GitHub and all the data therein.

\glspl{its} are conceptually simple tools which enable updating the properties of various issues that can be linked to each other to denote interdependencies \cite{Lueders:REJ:2023}.
Two teams could use the same \gls{its}, but due to the high configurability of these systems, the data and workflows within the respective \glspl{its} of these teams could look entirely different.
Teams can define custom issue types, custom properties, custom links, and custom workflows.
This high flexibility of \glspl{its} and the different backgrounds of stakeholders involved can lead to redundant~\cite{Zhang:TSEM:2023}, ambiguous~\cite{Ferrari:ASE:2019}, or conflicting issue  data~\cite{Zimmermann:TSE:2010}.
Possible consequences include poor issue report quality, intricate issue triage, delayed issue resolution, and communication difficulties ~\cite{Baysal:FSE:2014,Cavalcanti:JSEP:2014,Zou:TSE:2020,Zhang:SCIS:2015}.
Moreover, \glspl{its} can accumulate thousands or even millions of issues throughout the lifetime of a project~\cite{Anvik:OOPSLA:2005,Regnell:REFSQ:2008}, which may challenge stakeholders to locate and update the information they need, eventually compromising their productivity~\cite{Ho:GROUP:2001,Fucci:ESEM:2018}.

While prominent research on issue tracking can be found as early as the 2000s, including the seminal work on ``what makes a good bug report''~\cite{Bettenburg:FSE:2008, Zimmermann:TSE:2010}, user studies investigating the problems practitioners actually face within \glspl{its} remain rather rare.
Researchers have proposed numerous automation approaches, for example to detect duplicate issues~\cite{Zhang:TSEM:2023,He:ICPC:2020,Deshmukh:ICSME:2017}, facilitate issue creation~\cite{Heck:IWPSE:2013}, assign issues to developers \cite{Stanik:ICSME:2018}, predict the priority or severity~\cite{Izadi:EMSE:2022,Catolino:JSS:2019,Baysal:ICPC:2009}, or detect missing dependencies between issues~\cite{Lueders:RE:2022}.
However, comprehensive studies on the usefulness on those approaches and, first and foremost, on how developers and managers actually \textit{experience and use \glspl{its}} remain rare.
Particularly, little is known about the typical problems stakeholders encounter within \glspl{its}.
We think that this is surprising since understanding the challenges of \glspl{its} users is key for designing effective and efficient \glspl{its} approaches. 

Only recently, researchers started discussing specific issue tracking ``smells'', including how to detect and avoid them, such as community smells~\cite{Tamburri:IEEESoftware:2016}, agile smells~\cite{Telemaco:IEEEAccess:2020}, and bug tracking process smells~\cite{Qamar:SEAA:2021,Qamar:IST:2022}.
The concept of a ``smell'' is borrowed from ``code smells'' in which something is ``smelly'' if it \textit{may} lead to a problem, but \textit{not necessarily}.
For example, if issue properties like the assignee, priority, or severity are left blank for many issues \cite{Qamar:SEAA:2021,Qamar:IST:2022}.
This might result in problems with planning and prioritising the work, as well as usability problems for stakeholders trying to retrieve relevant issues (e.g. with a specific priority or assignee).
While approaches on how to detect and potentially avoid \gls{its} smells are starting to emerge, it remains, again, crucial to understand practitioners' perception of the smells, and particularity, when and why smells are considered problematic in practice.

This work takes a first step to fill the gap between a) understanding the needs and challenges of \gls{its} users and b) designing effective automation and assistance approaches in \glspl{its}.
Our objective is to gather an in-depth understanding of recurrent \textit{problems} software practitioners face when working with modern \glspl{its}.
In particular, we aim to understand how issue \textit{smells} discussed in recent literature impact their work, and when they might actually lead to problems. 
We also gather feedback on how smell \textit{tool support} in practice should look like.
Our research questions are as follows:
\begin{description}
    \item[RQ1.] What \textit{problems} do software practitioners usually encounter when using \glspl{its}?
    \item[RQ2.] How do practitioners perceive \gls{its} \textit{smells} discussed in literature: do they occur and are they problematic?
    \item[RQ3.] What do software practitioners think about \textit{tool solutions} to detect and handle \gls{its} smells?
\end{description}

To answer the research questions, we conducted an interview study with 26 practitioners from 19 companies (Section~\ref{sec:method}).
Our interviewees were experienced practitioners, who work as developers, managers, and product owners, and are knowledgeable about 18 different \glspl{its}.
We applied \gls{ta} to the interview notes to structure their reported problems with \glspl{its}, and we applied closed-coding analysis to their perspectives on smells reported in the \gls{se} literature.
The results of RQ1 reveal 14 common \gls{its} problems within three major themes: \gls{its} workflow problems, \gls{its} information problems, and \gls{its} organisational problems (Section~\ref{sec:problems}).
The results of RQ2 include in-depth insights on which \gls{its} smells occur and how problematic they are to practitioners, with specific context factors (Section~\ref{sec:smells}).
The results of RQ3 describe describe the participant feedback on potential tooling solutions to manage the smells (Section~\ref{sec:tooling}).
We present related work and compare with our findings (Section~\ref{sec:rel-work}).
We then discuss how our findings can guide researchers and tool vendors to focus future \gls{its} improvement approaches (Section~\ref{sec:disc}).
We also discuss ways in which \gls{its} users can adapt their workflows and automations to better manage the smells and problems identified in this study.
Finally, we discuss potential threats to the validity of our study (Section~\ref{sec:threats}) and conclude the work (Section~\ref{sec:conc}).

The contributions of this article are threefold.
First, our work offers a structured and detail-rich perspective on problems practitioners face when using modern \glspl{its}.
This is a prerequisite to design \gls{its} approaches that are effective and usable. 
Second, our work provides a practitioner-grounded perspective on the emerging research area of smells in \glspl{its}, detailing which smells occur in practice (i.e.~their prevalence) and which are problematic (i.e.~their severity).
Third, our work provides researchers and practitioners detailed insights into 13 specific context factors which affect smell occurrence and when smells lead to problems.
These context factors (and our detailed insights) can be leveraged to further investigate and improve \glspl{its}.

\section{Research Methodology}\label{sec:method}

We first reviewed literature to gather a non-exhaustive list of \gls{its} smells.
Then, we conducted 26 in-depth interviews with experienced industry participants.
Finally, we performed \gls{ta} and closed-labelling on the interview notes to form our findings.

\subsection{Issue Tracking Smells}
\label{sec:issue_tracking_smells}

\colorlet{customgrey}{gray!100}

\newcommand\tablesmell[2]{%
\renewcommand{\arraystretch}{0.1}%
\begin{tabular}[t]{@{}p{10cm}@{}}#1\\\setstretch{0}{\color{customgrey}\footnotesize\quad#2}\\\end{tabular}%
}

\begin{table}[ht]
\small
\centering
\renewcommand{\arraystretch}{0.8}     %
\caption{List of \gls{its} smells discussed in the interviews.}%
\label{tab:smells}%
\begin{tabular}[t]{@{} llp{10cm}l @{}}
    \toprule
     & \textbf{ID}  & \textbf{Smell} {\color{customgrey}\footnotesize{(and description underneath)}} & \textbf{Source} \\
    \midrule
    \parbox[t]{1mm}{\multirow{22}{*}{\rotatebox[origin=c]{90}{\textbf{Issue Property}}}}
        & S1.1  & \tablesmell{No or short (non-informative) description}{The ``description'' property is blank, or too short and non-informative.} & \cite{Prediger_2023_MSc} Ch.3 \\
        & S1.2  &  \tablesmell{Description too long}{The ``description'' property is too long.} & \cite{Lueders_2023_PhDThesis} Ch.4 \\
        & S1.3  &  \tablesmell{Issues are missing properties}{Properties such as assignee, priority, or severity are left blank on issues.} & \cite{Qamar:SEAA:2021,Qamar:IST:2022} \\
        & S1.4  &  \tablesmell{Issues are assigned to a team}{The value of the ``assignee'' property is a team, not an individual.} & \cite{Qamar:SEAA:2021,Qamar:IST:2022} \\
        & S1.5  &  \tablesmell{Often switching properties}{Properties such as status and assignee are switched back and forth between the same values.} & \cite{Halverson:CSCW:2006,Aranda:ICSE:2009,Prediger_2023_MSc} \\
        & S1.6  &  \tablesmell{Too many issue types}{There are too many options for ``issue type''.} & \cite{Kallis_2022_NLBSE} \\
        & S1.7  &  \tablesmell{No link to commit}{A resolving code commit is referenced in an issue, but there is no link to the commit in the issue.} & \cite{Qamar:SEAA:2021,Qamar:IST:2022} \\
        & S1.8  &  \tablesmell{Non-assignee resolved issue}{The person who marked the issue as ``resolved'' is not the person in the ``assignee'' property.} & \cite{Qamar:SEAA:2021,Qamar:IST:2022} \\
        & S1.9  &  \tablesmell{Ignored issue or delayed for too long}{An issue has been open for a long time, either delayed or completely ignored.} & \cite{Qamar:SEAA:2021,Qamar:IST:2022,Aranda:ICSE:2009} \\
        & S1.10 &  \tablesmell{No comments or too many comments}{The comment section of the issue has either no activity or too much activity.} & \cite{Qamar:SEAA:2021,Qamar:IST:2022,Tamburri:IEEESoftware:2016} \\
        & S1.11 &  \tablesmell{Toxic discussions}{The commend section of the issue has an unproductive and harmful discussion.} & \cite{SayagoHeredia_2025_JSEP} \\
        & S1.12 &  \tablesmell{Properties discussed in comments but not updated in issue}{In the comments, someone has referenced a property and the value it should be, but not updated it in the property. For example, ``This ticket is now closed'', but the status of the ticket remains open.} & \cite{Lueders_2023_PhDThesis} Ch.4 \\
    \midrule
    \parbox[t]{1mm}{\multirow{16}{*}{\rotatebox[origin=c]{90}{\textbf{Issue Linking}}}} 
        & S2.1  & \tablesmell{Issue without any links}{There are no links defined in the ``issue links'' field.} & \cite{Lueders_2023_PhDThesis} Ch.4 \\
        & S2.2  & \tablesmell{Too many link types / Link types with overlapping meaning}{There are too many options for ``link type'' or the options have overlapping meaning.} & \cite{Lueders_2023_PhDThesis} Ch.4 \\
        & S2.3  & \tablesmell{Multiple links with differing types to the same issue}{Issue A and B are linked to each other more than once, with different link types.} & \cite{Lueders_2023_PhDThesis} Ch.4 \\
        & S2.4  & \tablesmell{Known link mentioned in comments but not documented}{Same as S1.12, but specific to links.} & \cite{Lueders_2023_PhDThesis} Ch.4 \\
        & S2.5  & \tablesmell{Mismatch between link types and properties}{Mismatch between an issue's property and linking. For example, an issue is closed as ``duplicate'' but no duplicate link is added.} & \cite{Lueders_2023_PhDThesis} Ch.4 \\
        & S2.6  & \tablesmell{Mismatch of linked issues priorities and statuses}{Issues are linked together, but their respective properties don't make sense. For example, A is a linked as a sub-task of B, but B is closed and A is still open.} & \cite{Raatikainen_2023_TSE} \\
        & S2.7  & \tablesmell{Epic without sub-issues or sub-issues without main-issue}{An Epic is defined, but there are no linked children issues, traditionally User Stories, and vice versa.} & \cite{Lueders_2023_PhDThesis} Ch.4 \\
        & S2.8  & \tablesmell{Circular dependencies between issues}{Issues A is linked as a dependency of B, and B is also linked as dependency of A.} & \cite{Lueders_2023_PhDThesis} Ch.4 \\
    \midrule
    \parbox[t]{1mm}{\multirow{19}{*}{\rotatebox[origin=c]{90}{\textbf{Development Process}}}}
        & S3.1  & \tablesmell{Unplanned work added during sprint}{After a sprint has begun, new issues are added to the sprint.} & \cite{Telemaco_2019_CIbSE,Telemaco:IEEEAccess:2020} \\
        & S3.2  & \tablesmell{Too many complex issues are assigned to the same sprint}{Too many difficult or complex issues have been added to a sprint.} & \cite{Telemaco_2019_CIbSE,Telemaco:IEEEAccess:2020} \\
        & S3.3  & \tablesmell{Issues are missing an estimate}{The ``estimate'' property is blank.} & \cite{Telemaco_2019_CIbSE,Telemaco:IEEEAccess:2020} \\
        & S3.4  & \tablesmell{Inconsistent or unclear estimate-scales}{The ``estimate'' property is unclear or different scales are used across issues.} & \cite{Prediger_2023_MSc} Ch.3 \\
        & S3.5  & \tablesmell{Sprint does not end at the scheduled time}{Instead of moving open issues to the next sprint, the sprint end date is extended.} & \cite{Eloranta_2013_APSEC,Eloranta:IST:2016,Telemaco_2019_CIbSE,Telemaco:IEEEAccess:2020} \\
        & S3.6  & \tablesmell{Sprints have different duration}{Sprints are not consistent lengths.} & \cite{Eloranta_2013_APSEC,Eloranta:IST:2016,Telemaco_2019_CIbSE,Telemaco:IEEEAccess:2020} \\
        & S3.7  & \tablesmell{Sprint length differs from recommended length}{The sprint length differs from the recommendations made by Agile.} & \cite{Eloranta_2013_APSEC,Eloranta:IST:2016} \\
        & S3.8  & \tablesmell{Sprint has to be (repeatedly) delayed}{The same sprint is delayed over and over.} & \cite{Telemaco_2019_CIbSE,Telemaco:IEEEAccess:2020} \\
        & S3.9  & \tablesmell{No acceptance criteria or too many}{The ``acceptable criteria'' property is empty, or has too many defined criteria.} & \cite{Prediger_2023_MSc} Ch.3 \\
        & S3.10 & \tablesmell{Acceptance criteria are not checked during testing}{The ``acceptable criteria'' property has values, but they are not checked during the testing phase.} & \cite{Prediger_2023_MSc} Ch.3 \\
        & S3.11 & \tablesmell{Acceptance criteria are changed during sprint}{The ``acceptable criteria'' property has values, but they are changed during the sprint.} & \cite{Prediger_2023_MSc} Ch.3 \\  %
    \bottomrule
\end{tabular}
\end{table}

Our goal was to produce a list of \gls{its} \textit{smells} that could be leveraged for understanding the perspective of practitioners.
We sought neither completeness nor representativeness of the smells chosen.
To the best of our knowledge, there does not exist a complete taxonomy of \gls{its} smells for us to select from.
Accordingly, we built a sufficiently large list of smells from the literature to cover different organisational and process aspects.
All smells involve \glspl{its} directly or activities performed inside an \gls{its}.
The final list of smells originates from 11 articles:
    Halverson et al.~\cite{Halverson:CSCW:2006},
    Aranda et al.~\cite{Aranda:ICSE:2009},
    Eloranta et al.~\cite{Eloranta_2013_APSEC,Eloranta:IST:2016},
    Tamburri et al.~\cite{Tamburri:IEEESoftware:2016},
    Telemaco et al.~\cite{Telemaco_2019_CIbSE,Telemaco:IEEEAccess:2020},
    Qamar et al.~\cite{Qamar:SEAA:2021,Qamar:IST:2022},
    Kallis et al.~\cite{Kallis_2022_NLBSE},
    Raatikainen et al.~\cite{Raatikainen_2023_TSE},
    Prediger~\cite{Prediger_2023_MSc},
    Lüders~\cite{Lueders_2023_PhDThesis} and
    Sayago-Heredia et al.~\cite{SayagoHeredia_2025_JSEP}.
We list the smells in Table~\ref{tab:smells}, grouped into three high-level categories: Issue Property, Issue Linking, and Development Process.
Issue Property smells directly involve issue fields such as the Summary, Description, and Status.
Issue Linking smells involve the specific field ``Link'', capturing the full complexity of the field and how it might be a problem when used incorrectly.
Development Process smells involve processes conducted within \glspl{its}, where the \gls{its} is concerned with the configuration and management of that process.
For example, we consider S3.6 (``Sprints have different duration'') to be an \gls{its} smell because this is configured and controlled directly within the \gls{its}.
Contrastingly, we are not concerned with which flavour of Agile they are using, since that is a decision made outside the \gls{its}.

\subsection{Interview Participants}

\newcommand\diag[2][0mm]{\hspace{#1}\adjustbox{angle=45, lap=\width-1em}{#2}}
\newcommand\ddiag[2]{\diag{#1}\diag[0.5mm]{#2}}
\newcommand\its[1]{#1}
\newcommand\oit[1]{\color{gray}\small(#1)}

\begingroup
\renewcommand{\arraystretch}{1.0}
\begin{table}[ht]
\centering
\caption{Overview of study participants.}
\label{tab:participants}
\begin{tabular}{@{} r lll lll llll @{}}
\toprule
\textbf{ID} & \textbf{Role} & \textbf{YoE} & \textbf{G} & \textbf{ITS}~\oit{other previously used ITSs} & \textbf{ITS~Size} & \textbf{Industry} & \textbf{CS} & \textbf{Country} \\
\midrule
\rc \textbf{P01} & Product Owner & 7   & male & \its{Jira}                            & 10k-100k & Automotive  & L & Germany \\
    \textbf{P02} & Developer & 4   & male & \its{Trac}                                & 10k      & Engineering & M & Germany \\
\rc \textbf{P03} & Developer & 10  & male & \its{Trac}                                & 10k      & Engineering & M & Germany \\
    \textbf{P04} & Manager & 4   & male & \its{Jira}                                  & 10k-100k & Maritime    & M & Germany \\
\rc \textbf{P05} & Manager & 6.5 & male & \its{Jira} \oit{Asana}                      & 100k-1m  & Database    & L & Germany \\
    \textbf{P06} & Developer & 6   & male & \its{Jira} \oit{Trello}                   & 100k-1m  & Energy      & M & Germany \\
\rc \textbf{P07} & Developer & 24  & male & \its{Jira} \oit{Watson}                   & 1m+      & Media       & L & Germany \\
    \textbf{P08} & Product Owner & 5   & male & \its{Jira}                            & 100k-1m  & Automotive  & L & Germany \\
\rc \textbf{P09} & Developer & 10  & male & \its{Jira} \oit{Mantis, Watson, Trac}     & 1m+      & Media       & L & Germany \\
    \textbf{P10} & Manager & 5.5 & male & \its{Trac}                                  & 10k      & Engineering & M & Germany \\
\rc \textbf{P11} & Developer & 8   & female & \its{Jira} \oit{Custom}                 & 10k-100k & Consulting  & M & Germany \\
    \textbf{P12} & Manager & 11  & male & \its{Jira} \oit{Brainstorm, Watson}         & 1m+      & Media       & L & Germany \\
\rc \textbf{P13} & Developer & 3.5 & male & \its{Jira} \oit{GitHub, Bugzilla}         & 1k-10k   & Engineering & L & Germany \\
    \textbf{P14} & Product Owner & 6   & male & \its{GitLab} \oit{GitHub, Jira}       & 100-1k   & Engineering & S & Canada \\
\rc \textbf{P15} & Developer & 25  & male & \its{RedMine} \oit{GitLab}                & 1k-10k   & Research    & L & Germany \\
    \textbf{P16} & Developer & 1.5 & female & \its{GitLab} \oit{GitHub}               & 100-1k   & Engineering & S & Canada \\
\rc \textbf{P17} & Developer & 15  & male & \its{GitHub, GitLab} \oit{Jira, Savanna}  & 100-1k   & Research    & L & Germany \\
    \textbf{P18} & Developer & 3.5 & male & \its{ServiceNow} \oit{Jira, SpiraTest}    & 1k-10k   & Technology  & L & Canada \\
\rc \textbf{P19} & Developer & 16  & male & \its{GitLab} \oit{RedMine, Jira, GitHub}  & 10k-100k & Research    & L & Germany \\
    \textbf{P20} & Developer & 9   & male & \its{GitLab, GitHub} \oit{Jira, BaseCamp} & 1k-10k   & Software    & M & Poland \\
\rc \textbf{P21} & Developer & 15  & male & \its{Mantis} \oit{Jira, Trello, GitHub}   & 100-1k   & Medical     & L & Germany \\
    \textbf{P22} & Developer & 4   & male & \its{Jira} \oit{Miro, Trello, GitHub}     & 10-100k  & Consulting  & L & Germany \\
\rc \textbf{P23} & Developer & 23  & male & \its{Custom, GitHub}                      & 10-100k  & Research    & L & Germany \\
    \textbf{P24} & Manager & 15  & male & \its{Jira, Azure}                           & 10-100k  & Consulting  & L & Germany \\
\rc \textbf{P25} & Manager & 17  & male & \its{Jira}                                  & 1k-10k   & Medical     & L & Germany \\
    \textbf{P26} & Product Owner & 3.5 & female & \its{Jira, Azure}                   & 1k-10k   & Engineering & L & Germany \\
\bottomrule
\multicolumn{9}{l}{\footnotesize \makecell[l]{Study participants by
    Unique \textbf{ID}, \textbf{Role}, \textbf{Y}ears \textbf{o}f \textbf{E}xperience, \textbf{G}ender, current \textbf{ITS} and (other previously used ITSs), \textbf{ITS Size}, \textbf{Industry}, \\ \textbf{C}ompany~\textbf{S}ize~\cite{Eurostats_Enterprise_Size}, and \textbf{Country}.}} \\
\end{tabular}
\end{table}
\endgroup

To gather a broad and rich understanding of \gls{its} problems, smells, and contextual factors in practice, we sought a \textit{diverse} set of participants along five primary dimensions: participant role, years of experience, types of \glspl{its}, company size, and industries.
A representative sample was not sought, as we did not aim to generalise our qualitative observations.
We contacted \gls{se} companies in our network (mostly based in Germany) and asked for recommendations for experienced employees who would fit our sampling scheme.
After a first round of 13 interviews, we sought additional participants who were more diverse than the first, with countries other than Germany, more female participants, participants with more years of experience, participants in larger companies, and participants who used different \glspl{its}.
This additional round of 13 interviews led to a sample we deemed diverse enough to represent a breadth of experiences using \glspl{its}.

All participants worked at \gls{se} companies and had experience using \glspl{its}, with some participants having 10+ years of experience using \glspl{its}.
In total, we interviewed 26 practitioners working in Germany, Canada, and Poland, listed in Table~\ref{tab:participants}.
Our participants had a range of work experience from 1.5--25 years (median 7 years).
They also held various roles such as Developer, Manager, and Product Owner.
Consistent across all of them was the regular task of developing software.
All participants used an \gls{its} in their current position and had a range of experience using different \glspl{its} throughout their careers.
The majority (14/26) primarily used Jira, with six more participants having experience with Jira.
Others used GitHub, GitLab, Trac, Azure, Bugzilla, Trello, Asana, Mantis, SpiraTest, RedMine, BaseCamp, Miro, Watson, Brainstorm, Savanna, ServiceNow, and Custom \glspl{its}.
The size of their \glspl{its} ranged from just a few hundred to over a million issues.
The company sizes ranged from tens to thousands of employees, covering different industries including automotive, medical, consulting, and energy.

\subsection{Interview Procedure}\label{sec:interview_procedure}

We conducted one-hour semi-structured interviews over video calls, guided by an interview protocol.
To encourage an open dialogue regarding their true thoughts on the quality and direction of their company's \gls{its} practices, we refrained from recording the interviews.  %
However, to minimise observer bias, \textit{two interviewers} were present in each interview session: one primary and one secondary.
The primary interviewer (always the second author) led the interview session, asked the questions, and took notes on key responses as far as possible.
The secondary interviewer, focusing on taking detailed notes, was one of the three remaining authors or the postgraduate assistant listed in the acknowledgements.
When needed, both interviewers could ask additional, follow-up, or clarification questions.
The two interviewers met within {\mytilde}24 hours of the interview for 15--60 minutes to discuss results and align the notes.
15 interviews were conducted in English and 11 in German.\footnote{The language was decided based on the preference of the participant. All German interviews were conducted by native German speakers who are also C2 fluent in English, who later translated their notes into English.}

Each interview consisted of three main parts: problems related to using \glspl{its} (RQ1), perceptions on \gls{its} smells (RQ2), and opinions on tooling to manage these smells (RQ3).
We only introduced key concepts when they were necessary to avoid biasing their answers to earlier questions.
We started with a short welcome session ({\mytilde}10 minutes) where we introduced ourselves and the study and asked about their role, experience, and work context.
For \textbf{RQ1}, we asked the interviewees about challenges they encountered during their work with \glspl{its}.
For \textbf{RQ2}, we introduced the participants to our \gls{its} smells and asked them about their experiences with these smells. %
In particular, we asked about the \textit{occurrence} of these smells (or not), as well as their opinions on how \textit{problematic} each smell is (or not).
For \textbf{RQ3}, we sought to gather insights about the potential usefulness of tools to manage the smells.
We first asked about tooling in general to collect ideas without the bias of a specific tool or feature.
We then presented four screenshots\footnote{The screenshots are available in our replication package~\cite{ReplicationPackage}, embedded in our study protocol slide deck that we presented to our participants.} to guide the discussion and stimulate in-depth answers: (1) a configuration screen for the detection of individual smells, (2) a dashboard summarising the detected smells across the \gls{its}, (3) a detailed view showing detected smells for the currently viewed issue, and (4) an issue-link graph visualisation highlighting link smells.
We list the full interview script in Table~\ref{tab:questions} and in our replication package~\cite{ReplicationPackage}

\newcommand\subheader[1]{\normalsize{\textbf{#1}}}

\begin{table}[ht]
\small
\centering
\caption{Interview script.}%
\label{tab:questions}%
\begin{tabular}[t]{@{}p{8mm}p{13cm}@{}}
    \toprule
    \subheader{ID} & \subheader{Question Section} and question \\
    \midrule
        \subheader{SA}      & \subheader{Opening and Problems} \\
    \midrule
        SA1     & \textbf{Work Experience and ITS Usage} \\
        SA1.01  & What is your current role? \\
        SA1.02  & How long have you been at your company? How long have you worked in your current role? What other roles have you fulfilled? \\
        SA1.03  & What is your current/previous experience with issue trackers? (name the tools, any comments) \\
        SA1.04  & Can you describe the use of the issue tracker in your organization? What are common tasks, size, popularity, frequency, workflows/processes? Types of issues? Other special things? \\
        SA1.05  & Does your company/team use issue linking? How? Why? Types of links? \\
    \midrule
        SA2     & \textbf{Problems with Issue Tracking} \\
        SA2.01  & What issue tracking/issue management problems have you observed in your projects? Specifically related to adhering/supporting workflows or practices, updating/using issue properties and information, updating/using issue links. \\
        SA2.02  & What makes these problems problematic? \\
    \midrule
        \subheader{SB}      & \subheader{Smells} \\
    \midrule
        SB1     & \textbf{Open Smell Question} \\
        SB1.01  & Do you encounter smells in your issue tracking system? Can you think (spontaneously) of some particular smells? (Ask them to follow up with example screenshots if possible). \\
    \midrule
        SB2     & \textbf{Issue Property Smells} \\
        SB2.01--SB2.12 & \textit{Participant is shown each of the 12 smells in this category and asked the following:} Have you observed any of these in your projects? Do you think these are (or might lead) to actual problems? Why? \\
    \midrule
        SB3     & \textbf{Issue Linking Smells} \\
        SB3.01--SB3.08 & \textit{Participant is shown each of the 8 smells in this category and asked the following:} Have you observed any of these in your projects? Do you think these are (or might lead) to actual problems? Why? \\
    \midrule
        SB4     & \textbf{Development Process Smells} \\
        SB4.01--SB4.11 & \textit{Participant is shown each of the 11 smells in this category and asked the following:} Have you observed any of these in your projects? Do you think these are (or might lead) to actual problems? Why? \\
    \midrule
        \subheader{SC}      & \subheader{Tool Support} \\
    \midrule
        SC1     & \textbf{General Questions} \\
        SC1.01  & Would a tool to detect and manage smells make sense? Why? What would be the main requirements? \\
        SC1.02  & Would a tool to detect and manage (missing) links make sense? What would be the main requirements? \\
    \midrule
        SC2     & \textbf{Specific Screens / Features} \\
                & \textit{For each screen below, ask for feedback: Would it be helpful or not? Do these features address some of the problems you face? Why? What problems might occur with the tool? Would you prefer a specific link type prediction or a general prediction where you must determine the link type?} \\
        SC2.01  & List of \gls{its} smells with configuration \& activate/deactivate functionality \\
        SC2.02  & \gls{its} Health Analytics \\
        SC2.03  & \gls{its} Link graph \\
        SC2.04  & \gls{its} Visualization paired with Link Prediction / Smell Analytics \\
    \bottomrule
\end{tabular}
\end{table}

\subsection{Analysis}\label{sec:analysis}

Following the interviews, we digitised the 3,660 recorded meeting notes for analysis, where each ``note'' represents a complete idea (usually a sentence).
Each note was given a unique ID based on the participant, interviewer, and interview question, allowing for what we call ``evidence tracing'' to support our analyses.
We then performed a \gls{ta} of the interview notes to produce our qualitative findings for RQ1, and a closed-coding analysis of the RQ2 responses to visualise the overall findings for the \gls{its} smells.

\textbf{Thematic Analysis of Interview Notes.}
The findings of the interviews were extracted using \gls{ta}~\cite{Braun_2006_QRP}.
The first three authors were involved in the entire analysis (phases 1--6), while the last author was only involved in phases 3--6.
In \textit{Phase 1}, each of the researchers independently read through the entire set of digitised notes from the meetings, to familiarise themselves with the data.
This phase involved a single ``iteration'' over the data, in which the researchers were allowed to take notes to ingest the information, but were not yet allowed to generate or assign codes.
The result of this phase was a thorough understanding of the data, such that we could begin generating codes without the possibility for new information to surprise us during the next phase.
In \textit{Phase 2}, each of the researchers independently read through the data again, but this time they recorded ``codes'' which they felt represented a recurring concept within the reported problems.
For example, ``Lack of Automation'' was recorded by one researchers in response to comments such as ``no automation at all'' (P02) and ``sub-tasks resolved don't automatically resolve tasks'' (P05).
For every generated code, the researcher also had to record the ID of the participant statement, such that we could trace backwards from the code to the statement.
The result of this phase was three sets of independently generated codes (one for each researcher), where each set contains evidence tracing.
These three sets of codes are available in our replication package~\cite{ReplicationPackage}, including the evidence tracing IDs, but the evidence itself has been removed in accordance with our agreement with the participants to not release their interview data (see Section~\ref{sec:interview_procedure} for more details).
In \textit{Phase 3}, the researchers finally met to merge their codes into a single cohesive set.
We combined duplicates, aligned synonyms, and resolved disagreements about the nature of participant statements.
Disagreements were handled on a case-by-case basis, with open discussion and a unanimous result.
The result of this phase was a single set of codes, all of which have evidence tracing with combined evidence from all three researchers.
In \textit{Phase 4}, the researchers met to create an initial set of themes based on our shared understanding of the codes and their respective evidence.
This process was iterative in nature, involving four separate sessions where we met to group the codes, refine the meaning of themes, break apart large themes, combine smaller themes, and maintain cohesiveness across the entire thematic map.
The result of this phase was the full thematic map of grouped of themes, codes, and evidence.
In \textit{Phase 5}, we named and defined the themes, taking special care to avoid overlap in meaning and create easy to understand names.
The result of this phase (combined with phases 3 and 4) is a final set of grouped and named themes, codes, and evidence.
All data is available in our replication package~\cite{ReplicationPackage}.
In the final phase, \textit{Phase 6}, we selected representative examples (quotes) for this article, which you'll find in the results section below.

\textbf{Closed Coding of Smell Responses.}
To construct an overview of the participant feedback on \gls{its} smells, we conducted a closed-coding analysis of the RQ2 interview notes.
Closed coding is the process whereby researchers ``extract values for quantitative variables from qualitative data (often collected from observations or interviews) in order to perform some type of quantitative or statistical analysis''~\cite{Seaman_1999_TSE}.
In our case, we wanted a quantitative overview of the participant feedback on \gls{its} smells.
This was possible due to the categorical nature of the responses to our questions for each smell: ``have you observed this smell'' and ``do you think it is problematic''.\footnote{These are summarised; see our interview script in Table~\ref{tab:questions} or our replication package~\cite{ReplicationPackage} for original questions.}
The natural closed codes for these answers is one of the following: ``yes'', ``no'', ``depends'', or ``no answer''.
We applied the ``depends'' label when the participant explicitly stated that there are some situations in which the smell is problematic, and others where it is not (although the reason was not always stated).
We coded ``no answer'' when the participant had either nothing to say about the smell, or their response was not conclusive enough to be coded into one of the other three categories.
The first and third author first coded the responses separately, and then they met to resolve the disagreements.
We describe the coding agreement and results of this process in Section~\ref{sec:smells}.

\textbf{Replication Package.}
We created a replication package~\cite{ReplicationPackage} that includes all research artefacts discussed in this article, except for the digitised meeting notes.
Statements and knowledge about \glspl{its} can be quite critical (both for the business and the individuals), and so to protect the privacy of the interviewees, the digitised notes are not archived.
We do, however, share the remaining artefacts which have the IDs embedded in them, allowing for interpretation of the observations in context: you can trace analysis to the participant table (Table~\ref{tab:participants}) and relate to their anonymous details.

\section{RQ1. Common Problems}\label{sec:problems}
Here we describe the recurrent problems our participants observed in their daily work with \glspl{its}.
The themes and codes below are the result of \gls{ta} applied to the participant statements.
They are given in descending order of the number of participants mentioning each problem.
The participants had not yet been shown our list of smells; therefore, these categories represent an unbiased perspective on practitioner problems with \glspl{its}.

\subsection{ITS Workflow Problems}
Most participants (23/26) mentioned workflow problems within \glspl{its}.
Many mentioned the lack of fitting workflows (17) and the need for more workflow support (15), while others said that there is too much workflow bloat (14) and a lack of workflow enforcement (6).
This outlines a complex and nuanced trade-off within these systems between too much and too little support workflow options.

\textbf{Lack of a Fitting Workflow.}
Seventeen participants mentioned the problem of a Lack of a Fitting Workflow in their \glspl{its}.
They described limitations in issue fields and workflow options.
P26 said ``there are so many exceptions in real life; the process is valid for 90\% of use cases, but the 10\% of cases do not fit, so you have to cheat the process to make it work for these 10\%''.
One potential reason for the lack of fitting workflows is due to the ``pre-defined workflows'' that don't allow for flexibility, such as when ``a customer needs something quick'' (P04).
Another reason could be the desire for workflows to reflect the ideal way of doing things, while ignoring the way teams actually work.
As described by P08, \gls{its} workflows tend to be ``quite strict'', despite the fact that ''no team follows agile strictly''.
One consequence of a lack of fitting workflows is users are left with no choice by to ``cheat the process'' to complete their work (P26).
Another consequence is that users are ``hindered or inhibited from putting information into the \gls{its}'' (P17).
Lightweight \glspl{its} might not offer advanced customisation options and could potentially be more at risk for this problem.

\textbf{Lack of Workflow Support.}
Fifteen participants mentioned the problem of a Lack of Workflow Support in their \gls{its}.
Examples of desired automations include setting an issue from ``in progress'' to ``in review'' once a fixing commit was made, or closing a task once all sub-tasks were closed.
A potential reason for this problem is the initial creation of automated workflows.
P10 described their \gls{its} as having ``many options to customize, realization is rather difficult''.
Similarly, P17 mentioned that ``being able to configure Jira properly is a difficult task that requires advanced knowledge of development workflows and software development practices''.
The consequence of this lack of automation is ``lots of manual work'' (P03), which may then be done incorrectly or not at all.
This problem was described by a majority of developers (11/16) and half of managers (3/6).

\textbf{Workflow Bloat.}
Fourteen participants mentioned the problem of complex workflows within their \gls{its}, which some referred to as ``Workflow Bloat''.
This includes too many issue fields, required steps, and over-engineered workflows with numerous edge cases and excessive maintenance overhead.
P13 mentioned that there are ``too many link types, a whole drop-down menu''.
P17 reported a ``fine granularity of issue fields with a very granular workflow'', which sometimes leads to ``fitting yourself into the workflow, instead of just doing the work''.
Another challenge is the amount of information that must be entered while creating an issue report, particularly when the required information is not known at the issue creation time, or when the \gls{its} is configured with overly strict rules.
This can be time-consuming and result in additional manual work.
P02 and P20 mentioned examples where one small bug fix needed several hours of documentation.
The consequences of Workflow Bloat include confusion regarding what needs to be done and uncertainty regarding the importance of the required information.
P25 mentioned that ``with all of the possibilities, you can lose the focus on what you actually want to achieve''.
One reason for Workflow Bloat is older \glspl{its}, given the possibility for outdated fields and processes.
Another is when many teams with different internal workflows all contribute to the same \gls{its} configuration.
A more powerful \gls{its} can also be a reason, given the possibility for more complex workflows.
As P05 mentioned, ``Jira tries to mimic a perfect world where everything is known''.

\textbf{Lack of Workflow Enforcement.}
Six participants mentioned the problem of a Lack of Workflow Enforcement in their \glspl{its}.
This problem includes a lack of enforcement of required fields, accepted norms of \gls{its} usage, and workflow steps that are assumed to be consistent.
About half of the participants (more managers than developers) mentioned that \glspl{its} can allow ``too much freedom'', resulting in erroneous user input, such as misusing specific properties.
P06 mentioned that the ``environment property is a large text field where sometimes the description is erroneously entered''.
This problem may be caused by insufficient \gls{its} configuration or a lack of feature support within the \gls{its}.
One consequence of a Lack of Workflow Enforcement includes missing information, such as ``a bunch of issues in the backlog without labels'' (P14) or skipping important statuses.

\textbf{Unclear Workflow.}
Five participants mentioned the problem of Unclear Workflow in their \gls{its}.
This includes issue fields with unclear, vague, or differing definitions.
P05 mentioned ``Jira tries to establish a naming standard (blocked, resolved, etc.), but what does `resolved' actually mean?''.
One potential cause for Unclear Workflow is that clarifications and established conventions cannot be represented in the \gls{its} itself.
P06 stated that their ``team has a definition of `done'/`ready' that cannot be made transparent in Jira''.
This problem may lead to users spending more time than intended on the issue documentation, or alternatively skipping steps because of decision fatigue.

\subsection{ITS Information Problems}
Participants described difficulties retrieving information from the \gls{its} and maintaining a general understanding of the overall \gls{its} state.

\textbf{Missing Issue Information.}
Eighteen participants mentioned problems related to the accuracy and completeness of information in their \glspl{its}.
This includes information that is incorrect, irrelevant, or missing, as well as vague definitions and insufficient details in issue descriptions (partly confirming results by Zimmerman et al.~\cite{Zimmermann:TSE:2010}).
These problems were more often mentioned by developers (12/16) than managers (3/6). %
P07 said that the ``main problem is the people who did not enter enough information''.
The primary cause for this appears to be friction during the information entering process.
Users may be unsure about how to fill in certain fields, so they revert to skipping the field or filling in incorrect information.
An additional cause for this problem is that information may be ``stored inside other communication tools such as Mattermost'' (P19) without being added to the ticket.
Without the correct information, resolving issues takes a longer time due to required clarifications, or may not happen at all.
One reason for Missing Issue Information appears to be when \glspl{its} are used by many untrained or external people.

\textbf{Ineffective Search.}
Ten participants mentioned the problem of ineffective search within their \glspl{its}.
They described difficulties searching for specific issues, as well as classes of issues such as open Epics associated with a certain project, due to limitations of the search features in the \gls{its}.
Even with sophisticated filtering, the phrases required to locate the issues were not easily obtainable.
P06 mentioned that ``you have to search for the exact title to find a ticket''.
Similarly, P09 found that ``there are many synonyms, which makes it harder to find something''. 
This would result in the search returning no issues, or too many issues related to non-important aspects of the search term (confirming findings from Heck and Zaidman~\cite{Heck:IWPSE:2013}).
Difficulties in locating relevant existing issues lead to a less connected \gls{its}, as well as duplicate issues.
Ineffective Search is particularly severe in larger \glspl{its}.

\textbf{Issue Overload.}
Eight participants mentioned the problem of too many issues (both open and closed) within their \glspl{its}.
P15 stated that ``having so many issues can be a problem. We are not sure what to do with all of these issues''.
Similarly, P18 described getting ``lost navigating around Jira''.
The consequences of this smell include a lack of awareness of existing issues, leading to duplicates and overlapping issues, missed links, and issues being forgotten.
One such reason for this problem is older \glspl{its}, which have accumulated more issues over time.
Issue overload appears to affect managers more than developers.

\textbf{Zombie Issues.}
Eight participants mentioned the problem of inactive and abandoned issues within their \glspl{its}, which some referred to as ``Zombie Issues''.
This issue was mentioned mostly by developers (7/16) and not managers (1/6).
They described the problem as a lack of attention towards issues that were still open, but were waiting for some action to be taken.
Specific causes for this problem include when the reporter and assignee are the same person (accountable only to themselves), and low-priority issues (minimal pressure to act).
P07 said some issues get ignored, and that ``it happens to all developers'' due to a ``focus on other topics''.
Users who are aware of ignored issues may still decide to keep them open, as P15 described being ``reluctant to close issues because they are not urgent and not important''.
Zombie Issues may lead to a clogged backlog and duplicate issues (due to unawareness of existing issues).
P06 described that ``delayed tickets lead to a huge backlog of irrelevant tickets''.
The reasons for this problem include older and larger \glspl{its} (where more issues is the stressor), as well as multiple disjoint teams working loosely together (which leads to uncertainty about respective responsibilities).

\textbf{Lack of Comprehensive Overview.}
Eight participants mentioned the problem of a lack of a comprehensive overview.
They described difficulties gaining a holistic understanding of the current state of their \glspl{its}, as well as their own tasks and responsibilities.
P08 stated that ``it is hard to keep an overview of the huge backlog, and keep the projects separate from each other''.
They expressed a desire for a better grasp of the big picture and the dependencies between issues.
P10 mentioned that ``a tree would be nice to understand dependencies''.
Participants described the cause as missing features in their \gls{its}. 
While some \glspl{its} facilitate custom dashboards, these must be created by the users, who are themselves often not exactly sure what to visualise.
The consequence of a lack of an overview is the repeating need to manually find things, as well as the need to remember (or write down) exactly what to keep track of.
The desire for such a dashboard was mentioned by a higher percentage of managers (2/6) and product owners (2/4) than developers (4/16).

\subsection{ITS Organisational Problems}
Some participants described problems involving information, processes, and collaboration across organisational contexts.

\textbf{Divergent Tracking Needs.}
Eleven participants mentioned the problem of divergent needs in their \gls{its}, particularly across teams and projects.
This includes variances in issue fields and available properties, but also entire workflows supported by the \gls{its}.
As described by P17, there is a ``heterogeneous usage of Jira, even within the same project, which means no one is following the same workflow''.
Similarly, P06 explained that ``only Jira admins can edit workflows, so individual teams can't make adjustments as needed''.
A consequence of Divergent Needs is that \glspl{its} become too diverse with the number of available options and start exhibiting Workflow Bloat, thereby creating confusion as to the correct options for a given workflow.
This can lead to communication issues and a lack of understanding about the processes being used by different teams, including their own.
One reason for this problem is larger companies with many projects and teams, all using the same \gls{its} instance and configuration.

\textbf{Information Islands}
Ten participants mentioned the problem of multiple disjoint systems and missing integration in their \glspl{its}~\cite{Maalej:ASE:2009}, leading users to duplicate, split, or omit information across systems.
Such disjoint systems include Git, communication and documentation tools, and even other \glspl{its}.
As P20 mentioned, there are ``multiple sources of truth: Slack, Jira, etc.''
A common reported problem was the use of multiple \glspl{its} within the same organisation, with slightly different use cases (e.g.~one public and one private \gls{its}).
One cause of Information Islands is the lack of useful integration between systems that are used in parallel.
The consequences of this problem include duplicate and missing issues, as well as increased effort required to create, maintain and retrieve the information across systems.
As P15 stated, ``people have tried to implement processes between the two issue trackers'' at their organization.

\textbf{Lack of Mindset \& Discipline.}
Nine participants mentioned the problem of a Lack of Mindset \& Discipline within their \gls{its}.
P15 and P22 both mentioned that good discipline is needed.
P22 elaborated, saying that ``\glspl{its} must be used properly to be successful, this has nothing to do with the \gls{its} itself''.
This problem also includes things like waiting for people, difficulty choosing an estimate or due date, or inappropriate communication.
Managers are often responsible for organising issues and keeping them up-to-date, which can lead to developers creating issues with lower quality.
P12 stated that ``dates and estimates are very difficult for developers''.
As Meyer et al.~\cite{Meyer:FSE:2014} observed, developers feel productive when they close many or big tasks, whereas creating or organising issues feels unproductive.

\textbf{Scoping Issues is Hard.}
Nine participants mentioned that Scoping Issues is Hard in their \glspl{its}.
They described difficulties breaking apart and appropriately scoping large issues.
Unclear or under-defined issues are a potential cause for this problem.
Our participants described that \gls{its} users may struggle with uncertainty and ambiguity, resulting in issues lacking clarity in their scope or being scoped incorrectly, leading to ``issues that are too small or too large'' (P13).
As a consequence, users might incorrectly assume issues to be done and close them too early, leading to issues needing to be reopened.
As P14 pointed out: ``what is the definition of `done' for this work?''.
This problem was described by developers (7/16) and product owners (2/4).

\section{RQ2. Smells}
\label{sec:smells}

\afterpage{%
    \clearpage%
    \begin{figure}[ht]
        \begin{adjustbox}{addcode={\begin{minipage}{\width}}{%
            \caption{Participant responses on whether the smells occur, and whether they think they are problematic. Yellow: Agree, Red: Disagree, Blue: It Depends, Blank: No Response.
            Note that the question of ``occurrence'' does not warrant the ``depends'' code, since the participant either experienced the smell or not.%
            \Description{This figure shows two tables, where the rows are participants and the columns are the smells. The first table is the responses to the question ``do these smells occur'', and the second table is the responses to the question ``are these smells problematic''. The intersection of each row and column is a single square with a colour: yellow is agreement, red is disagreement, blue is it depends, and blank is no response.}
            \label{fig:smell_responses}}
            \end{minipage}},center}
            \includegraphics[width=.66\textwidth]{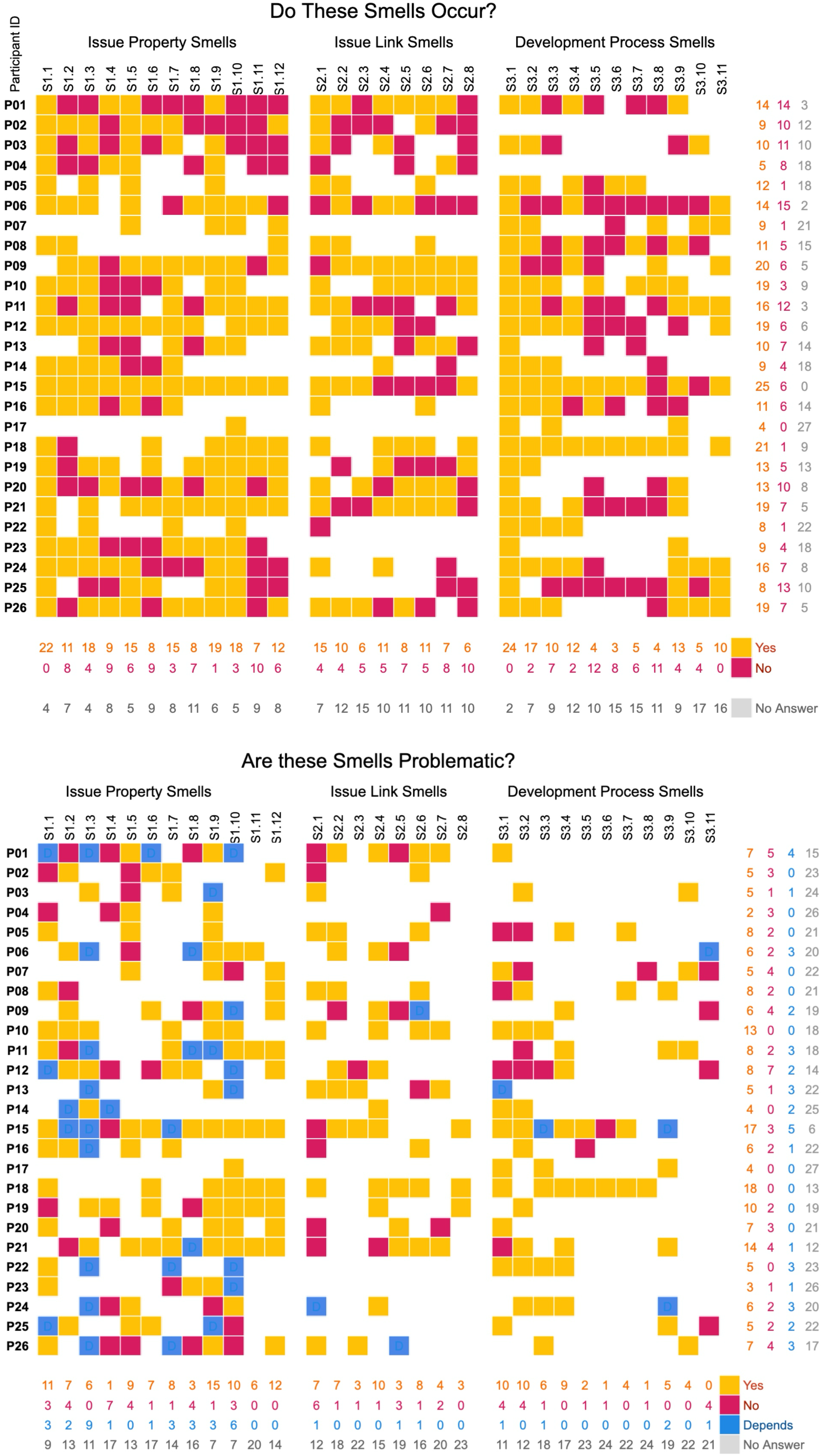}%
        \end{adjustbox}
    \end{figure}
    \clearpage%
}

\newcommand*{\contextBox}[1]{\tikz[baseline=(X.base)]\node [draw=black!50,fill=cyan!10,thick,rectangle,inner sep=2pt,rounded corners=3pt] (X) {#1};}
\newcommand*{\contextName}[1]{\contextBox{#1}}

We describe here the results of showing each participant our list of 31 \gls{its} smells and getting their feedback on 1) do these smells occur, and 2) are they problematic.
As described in Section~\ref{sec:analysis}, we used closed-coding analysis to convert the qualitative data into quantitative insights.
The initial coding of responses showed 74\% agreement ($\kappa = 0.61$), with a final agreement of 100\% after a correction and alignment session.
We visualise the closed-coded smell responses in Figure~\ref{fig:smell_responses}.
Each cell in the figure represents a response from one of the participants.
The columns represent each of the 31 smells, and the rows are the individual participants.
The bottom and right sides of the figure include the counts of responses for each smell and participant.

All the smells were noted as occurring by at least some participants, and most of the smells were also noted as problematic.
However, when asked about the smells being problematic, 98 of the 290 comments were either ``no'' or ``it depends''.
The 31 smells originate from 11 source articles (see Section~\ref{sec:issue_tracking_smells}) that suggest these smells exist and can lead to \textit{problematic} outcomes if not addressed.
Therefore, the participant responses of ``no'' and ``it depends'' (to whether the smells are problematic) are somewhat contradictory to the claims made in the 11 source articles.
We therefore focus our analysis on these responses.
This means, we combine the ``no'' and ``it depends'' responses into the generic concept of ``disagreements'' for the purpose of summarising and discussing the contradictory evidence we found in this study.

Finally, while discussing the smells, the participants sometimes mentioned specific context factors related to their experiences (or lack thereof) with the smells.
We highlight them (like \contextName{this}) and discuss them in the smell where they were most often mentioned.
We also summarise them below in Section~\ref{sec:context_summary}.

\subsection{Issue Property Smells}

Participants largely agreed that Issue Property Smells occur, but 28 of their responses said they are unproblematic, and 31 responses noted that ``it depends'' with some also mentioning specific context factors.

\textit{Issues are assigned to a team (S1.4)} received the most disagreement of the Issue Property smells (8/9 participants that answered).
Most of the reasons for disagreement mentioned the context factor of \contextName{intended outcome}, whereby the smell is actually the result of an intentional workflow, and thus not a smell for these participants.
P01 and P26 mentioned that issues are assigned to a team first, then to an individual.
P01 and P24 said that issues are assigned to teams to ``engage'' the teams, with the main assignment left to them to ``decide autonomously''.
P12 said that team assignments can be used for downstream activities such as ``in review'', whereby the team responsible for reviewing the issue will be assigned.
The context factor \contextName{\gls{its} configuration} was also common for this smell, whereby the smell does not occur because the \gls{its} is configured in such a way to prevent it entirely, or help keep it unproblematic.
For example, P11 mentioned that they have ``no team accounts, only individual people'', and thus this smell cannot occur.
Finally, participants mentioned the context factor \contextName{team size}, whereby the smell is irrelevant due to the size of the team.
In situations where ``a team consists of one person'' (P23) it is not possible to assign issues to a team.

\textit{Non-assignee resolved issue (S1.8)} received many disagreements (7/10 participants).
Participants mentioned the context factor \contextName{workflow design}, whereby their workflow prevents the smell from occurring in the first place, or prevents it from being problematic.
P06 mentioned that this smell ``happens and is resolved in sprint planning``, therefore it is ``not a problem'' for them.
One participant, P01, mentioned the context factor \contextName{supervisor involvement}, whereby it is not a smell because he---as the manager---is ``the only one responsible for closing all the tickets''.

\textit{Issues are missing properties (assignee, environment, priority, severity, ...) (S1.3)} also received many disagreements (9/15).
Participants again often mentioned the context factor \contextName{intended outcome} for this smell.
P15 and P16 both said that fields such as the assignee are assigned later, such as on the first day of the sprint.
P01 mentioned something similar, remarking that ``no assignee while the ticket is in progress would be considered bad''.
Participants also mentioned the context factor \contextName{\gls{its} configuration}.
P06 and P24 mentioned that some fields are enforced by the \gls{its}.
This means that they are so problematic if left empty, that the companies have put a technological barrier in place to prevent them from being blank.
Lastly, the most common context factor mentioned by participants for this smell is \contextName{issue property}, whereby the smell is problematic for some properties, but not for others.
P26 said that they ``differentiate important vs non-important properties``, giving the example that severity if an important field and therefore ``almost never missing``, whereas non-important fields such as Tags and Components are ``not so harmful if they are missing''.
P05 agrees with this sentiment, stating that there are ``properties nobody cares about that are often missing''.

\textit{No comments or too many comments (S1.10)} had 9/19 participants in disagreement, based primarily on the context factor \contextName{intended outcome}.
P01 said that he doesn't consider too many comments bad, as his ``team doesn't use comments that much''.
P02 mentioned that instead of comments, the ``information is added to the description''.
P06 said that ``no comments means everything is well'', and therefore an intended part of their workflow.

\textit{Description too long (S1.2)} was another smell with split opinions (6/13).
Participants mentioned the context factor \contextName{team preference}, whereby the participant's team simply chooses to work this way, thereby the smell is not problematic for them.
P08 said that they ``value longer and more complex descriptions'', and P11 said ``the more information the better''.
Participants also mentioned the context factor \contextName{issue type}, whereby the type of the issue affects whether the smell is problematic or not.
P14 said that ``too long without a purpose is a problem'', as bug reports need shorter descriptions, but ``sometimes epics need longer descriptions''.
P03 mentioned that long descriptions are not a problem, and in fact, short descriptions are ``sufficient for bug reports''.

\textit{No or short (non-informative) description (S1.1)} had 6/17 participants in disagreement.
The main reason participants said a short (or empty) description was not problematic was the context factor \contextName{issue complexity}, whereby how problematic the smell is depends on how complex the issue is.
P04 mentioned that it is often sufficient if the issue has a Parent issue, thereby offloading the complexity of the issue to the parent description.
This hierarchical structure of issues can lead to well-defined Epics, with children User Stories that just contain a title and no description.
P01 agrees with this sentiment and said that ``sometimes there is no need for a long description, and only the title is sufficient''.
Participants also mentioned the context factor \contextName{stakeholder role}, whereby the role of the stakeholder interacting with the issue affects how problematic this smell may be.
While an empty description of a user story may be sufficient for the customer or requirements engineer, P26 said ``it is harmful for the team that has to develop the user story''.

\textit{No link to commit (S1.7)} had 4/12 participants in disagreement.
Participants brought up the context factor \contextName{\gls{its} configuration} as a reason this smell does not occur.
P09 said that when the Jira issue is linked within a commit message, the associated Jira ticket is automatically updated to include a link to the commit.
This automation was also mentioned by P06, P11, and P20.
Participants also mentioned the context factor \contextName{issue type}.
P15 said that it depends on the issue type, since issues such as support requests do not have to be linked to a commit.

\textit{Often switching properties (status or assignee, \dots) (S1.5)} had 4/13 participants in disagreement.
The primary disagreement about this smell is based on the context factor \contextName{intended outcome}.
P02 said that it is ``part of their process to pass issues back and forth between assignee and reviewer; it is their quality assurance'', and hence often switching the assignee property.
This sentiment was also held by P06, who said that ``it is intended''.

\textit{Ignored issue/delayed for too long (S1.9)} only had 4/19 participants in disagreement.
Participants disagreed with the occurrence of this smell based on the context factor \contextName{workflow design}.
P02 said that they ``meet once a week'' to discuss the issues, so they do not get delayed too long.
P11 said that ``tickets in a sprint are not allowed to be in the same status for longer than a day'', so they cannot be ignored.

\subsection{Issue Link Smells}

Participants largely agreed that Issue Link Smells occur, but 15 of their responses disagreed that these smells are problematic, and 3 responses noted a dependent context factor.

\textit{Mismatch between link types and properties (S2.5)} had the most disagreement with 4/7 participants, but very few mentioned context factors.
One context factor mentioned is \contextName{workflow design}, whereby the smell cannot arise in the team's workflow because it is not relevant.
For example, P19 said that ``they don't use properties much''.

\textit{Issue without any links (S2.1)} had many disagreements with 7/14 participants.
The primary context factor mentioned for this smell is \contextName{\gls{its} configuration}.
P04 explained that ``tickets are pre-generated, so there is no need to link projects''.

\textit{Mismatch between linked issues regarding their status, due date, priorities, or estimates (S2.6)} had 2/10 participants in disagreement.
P09 said that can happen, but only the status fields being out of sync matters, not the resolution.
Both P09 and P13 mentioned that this problem is largely solved by external communication regarding the issue.

\subsection{Development Process Smells}

Participants largely agreed that Development Process Smells occur, but 16 of their responses disagreed that these smells are problematic, and 5 responses noted a dependent context factor.

\textit{Acceptance criteria are changed during sprint (S3.11)} had full disagreement with all 5 participants that responded.
The consensus among them was that it is more important to update the acceptance criteria, than to leave it outdated when you know it should be changed.
The context factor \contextName{issue dependency} was mentioned as a context factor for why this may be necessary.
As P25 stated, ``it is important to update the acceptance criteria if the change is needed''.

\textit{Unplanned work added during sprint (S3.1)} had 5/15 participants in disagreement.
The five participants who disagreed all mentioned that unplanned work was a normal part of their sprint operations.
P08 said that it ``happens every sprint'' and is ``normal operations''.
P21 added that this is an important topic, and it is not a problem ``as long as you get your work done''.
Some participants mentioned the context factor \contextName{customer priority}, whereby the smell can occur due to customer prioritisation override or skipping regular workflows.
P07 said that an ``important customer finds something and this gets priority''.
Similarly, P01 said that ``customer pressure'' caused the smell to occur, and described it as ``stressful, inefficient, and it puts a lot of pressure on the team''.

\textit{No acceptance criteria or too many (S3.9)} had 2/7 participants in disagreement.
P25 said that it ``depends on who is asking for the feature'', and noted that if any problem exists, it's more about no acceptance criteria than too many.
P24 said that it is not a problem since ``no acceptance criteria happens and are created after the fact''.
P21 mentioned the context factor \contextName{stakeholder role}, explaining that ``acceptance criteria are only for developers'' and further testing is managed by other teams.

\textit{Too many complex issues are assigned to the same sprint (S3.2)} had 4/14 participants in disagreement.
The primary reason for disagreement was the context factor \contextName{workflow design}.
P05 and P07 both said the smell is unproblematic because you can just delay issues to the next sprint without repercussions.
Additionally, \contextName{issue complexity} was mentioned as a context factor for when this smell may be problematic.
P11 stated that complex issue count is not enough, since it is not a problem as long as the complex issues are distributed between people correctly.

\textit{Issues are missing an estimate (S3.3)} had 2/8 participants in disagreement.
Multiple participants mentioned that the smell may not be problematic due to the context factor \contextName{intended outcome}.
P12 said that they don't do estimates because of their very complex code base, which always leads to wrong estimates anyway.
P15 mentioned that it depends on the sector you are working in, since not all companies have to agree on estimates ahead of time.
The context factor \contextName{workflow stage} was also mentioned.

\subsection{Summary of Context Factors}
\label{sec:context_summary}

\begin{table}[t!]
\centering
\small
\caption{Summary of Context Factors described by participants when discussing smells. Whether a smell occurs and is problematic depends on these context factors.}
\label{tab:context}
\begin{tabular}{l c p{0.48\textwidth} l @{}}
    \toprule
    \textbf{Context Factor} & \textbf{Count} & \textbf{Description} & \multicolumn{1}{l}{\textbf{Most Relevant Smells}} \\
    \midrule
    workflow design & 48 &
        Smell is caught early or prevented entirely as part of a standard workflow, such as regular issue review meetings &
            S1.8 S3.2 S1.9 \\
    intended outcome & 44 &
        Smell is an intended or expected outcome of a certain workflow &
            S1.10 S1.4 S1.5 S3.3 S1.3 \\
    \gls{its} configuration & 24 &
        Smell cannot occur due to technical constraints by the \gls{its}, such as limitations, automations, or mandatory fields &
            S1.4 S1.7 S1.3 S2.1 \\
    supervisor involvement & 11 &
        Smell is either mitigated or monitored by the participant's supervisor and thus not a concern for them &
            S1.1 S1.3 S1.8 \\
    team preference & 9 &
        Smell is considered acceptable or negligible by the team &
            S1.2 S1.6 S1.1 S1.5 \\
    workflow stage & 8 &
        Smell is considered problematic only during some workflow stages, such as within the current sprint &
           S3.3 S1.3 S1.9 S1.9  \\
    issue property & 8 &
        Smell is prevented or considered problematic for some issue properties, and acceptable for others &
            S1.3 S2.4 S2.5 S2.6 \\
    team size & 7 &
        Smell is considered more or less problematic due to the size of the participant's team &
            S1.4 S1.2 S1.3 S1.9 \\
    issue complexity & 6 &
        Smell is only problematic depending on the issue complexity &
            S1.1 S3.2 S1.2 S2.1 \\
    issue type & 5 &
        Smell is considered problematic only for some types of issues, such as bug reports or feature requests &
            S1.2 S1.7 S1.1 S1.6 \\
    issue dependency & 5 &
        Smell is considered problematic based on some other nuanced per-issue context, such as how it relates to other issues &
            S3.11 S3.9 S3.10 \\
    stakeholder role & 5 &
        Smell is considered problematic for some \gls{its} users but not others &
            S1.1 S3.9 S1.6 S1.7  \\
    customer priority & 4 &
        Smell can occur due to customer prioritisation overriding or skipping regular workflows &
            S3.1 S3.3 \\
    \bottomrule
\end{tabular}
\end{table}

The participants mentioned various context factors that contributed to not encountering certain smells, or not considering smells problematic.
We summarise these context factors in Table~\ref{tab:context}.
The first three context factors (accounting for 63\%) describe rather absolute conditions under which the smells do not occur or are not problematic.
Employing the context factors \contextName{workflow design} and \contextName{\gls{its} configuration} both prevent or mitigate the smell entirely, while the context factor \contextName{intended outcome} describes contexts where the smell is actually desired.
The last ten context factors are more nuanced in how they describe when, where, and to whom the smell occurs and is problematic.
The context factors \contextName{supervisor involvement}, \contextName{team preference}, \contextName{team size}, \contextName{stakeholder role}, and \contextName{customer priority} all describe contexts where it matters \textit{who} is involved.
The context factor \contextName{workflow stage} describes situations where it matters \textit{when} the smell occurs.
The context factors \contextName{issue property}, \contextName{issue complexity}, \contextName{issue type}, and \contextName{issue dependency} all describe contexts where it matters \textit{what} artefacts, in which condition, are involved.

\section{RQ3. Tooling for ITS Smells}\label{sec:tooling}

Overall, our study participants highlighted that tool support for handling \gls{its} smells should be easy-to-use, well-integrated into their workflows, and reliable. 
By meeting these requirements, they think a tool could help to improve the efficiency and effectiveness of issue tracking. 
We first asked participants about any \textit{general tooling feedback} they have, and afterwards we showed them the screenshots of four prototypes (smell configuration, dashboard, smells in issue views, and smells in graph visualisation) and gathered feedback on those general \glspl{its} features.

\subsection{General Tooling Feedback}

\textbf{Prevention over Detection.}
Nine participants (P03, P05, P08, P09, P13, P16, P17, P24, P26) said that instead of detecting the smells, preventing them from happening might be more useful. 
Otherwise, there would still be the overhead of fixing smells when they occur and lead to actual problems.
For example, P05 said that ``smells should be prevented during issue creation, for example: if you want to close an issue as duplicate, you get a pop-up if there is no link''.
This is in alignment with the common context factors \contextName{workflow design} and \contextName{\gls{its} configuration}, both of which help to prevent and minimise the impact of smells within \glspl{its}.
The participants noted that an essential requirement for adequate tool support in issue tracking is the direct integration into their \gls{its} workflow~\cite{Maalej:ASE:2009}.
This means the tool should be easy to use and not require too much overhead while creating or modifying issues; otherwise, it will not be used.

\textbf{Tooling Pitfalls to Avoid.}
Nine participants (P06, P07, P12, P14, P15, P21, P22, P24, P26) expressed concern that the user might receive too many notifications.
P07 stated the system ``should not spam too much; if the information is not too obvious then it would be very helpful''.
Notifications can be annoying, particularly outside the \gls{its}, for example with e-mails~\cite{Meyer:TSE:2021}.
Additionally, participants mentioned that the tool should not obstruct the workflow, as it creates more overhead.
P13 said that ``enforced processes are not good'' and P22 said ``it should not completely block your workflow''.
In this case, the context factors \contextName{workflow stage} and \contextName{workflow design} are relevant, since some smells are only relevant during certain parts of the workflow, or not at all.
Multiple participants voiced their scepticism toward automated processes (P07, P08, P12, P15, P18, P21, P23, P24).
They said that the results need to be transparent, interpretable, and manually reviewable.
Participants noted that some smell detection would likely be more useful for managers than developers.
In these cases, the context factors \contextName{supervisor involvement} and \contextName{stakeholder role} are relevant, as some smells may apply only to certain roles. 
P14 also cautioned that there might be psychological  impacts of such a tool, making users feel bad instead of helping them improve. 

\textbf{Smell Detection Tooling to Support Meetings.}
Seven participants (P01, P08, P11, P13, P14, P20, P25) mentioned that smell detection tooling could be used to support the quality of their \gls{its} in meetings such as triage and sprint planning.
Multiple participants mentioned that triage meetings are essential to organising \glspl{its} and ensuring that issue reports have the necessary information for developers to act on them.
These meetings are already a way to ensure that issues in the sprint adhere to the expected quality standard and can be resolved.
Participants mentioned that smell detection could support these meetings by highlighting potential problems.
P14 also mentioned the potential to support retro meetings: ``it would be nice to look at a dashboard and say we were 50\% worse in this smell''.

\subsection{Feedback to Specific Smell Handling Features}

\textbf{(1) Configuration.}
We showed the participants a screenshot of a smell configuration page, containing a toggle for each smell (and some had a configurable value).
Almost all participants said that a configuration is a key requirement. 
A smell management tool ``must be customisable because some smells would not work or would constantly be detected'' (P03).
Our list of identified context factors in Table~\ref{tab:context} outlines different reasons why smells are not relevant in certain situations.
For example, the context factor \contextName{team preference} explains that some smells are informally considered acceptable of negligible by certain teams, and therefore they will want to configure them to be ``off''.
P10 mentioned that the tool could be misused and hinder more work than help if it is not configured properly.
P13 cautioned that ``deactivating smells is a nice feature, but they need to understand and document why smells are deactivated''.
P26 also added that a \textit{default} configuration is needed.
Finally, P05 mentioned that they would like ``syntax to write their own smells''.

\textbf{(2) Dashboard.}
We showed the participants a dashboard presenting the overall smelliness (i.e.~``health'') of the \gls{its}.
Multiple participants commented that a dashboard has more value for managers, with little value for developers.
This is particularly true for situations where the context factor \contextName{supervisor involvement} applies, since the supervisor should maintain an awareness of what is going on.
It was also noted that it should not be the absolute values but the trends over time.
P21 said it ``would be nice to see the change over time with the health, so it's not only the static view that is important, but the change over time''.
P24 said that tools like this should be used with a goal in mind because ``optimising for green isn't always good'' and that the metrics shouldn't become the targets as they can ``distract from what the customer wants''.

\textbf{(3) Smells in Issue View.}
We showed the participants an issue view with integrated feedback on detected smells for this individual issue.
Developers in our sample preferred smell detection for individual issues more than in an \gls{its} dashboard.
P17 said that ``the information needs to be pre-digested into actions being requested of the developers; they can then decide to do it, or not'' and P02 said ``but they don't need the stats, just the lists of smells and where/how to fix it''.
This is particularly true in situations with the context factors \contextName{intended outcome}, \contextName{workflow stage}, and \contextName{issue property}, since developers have to assess the relevance of these factors in deciding on the severity of the smell.

\textbf{(4) Smells in Graph Visualisation.}
We showed the participants a graph view of interlinked issues, including which link smells each issue has.
Participants mentioned that lists may be sufficient and visualisation is not needed for managing smells and links.
Visualisations may be preferred at higher levels (e.g.,~epics) but may be overkill at lower levels (e.g.,~subtasks).
P22 said that this also depends on the usage of links: ``depends how you work with it, what kind of project you are working in, and if a team uses a lot of issue links: then this would be helpful''.
In these cases, the context factors \contextName{issue type} and \contextName{issue complexity} are important, since only certain issue types such as ``Epic'' are affected, and the more complex an issue is the more likely the links could benefit from a visualisation.

\section{Related Work}\label{sec:rel-work}

While interview studies of \gls{its} usage are rare, there are numerous studies about the quality of issue reports (particularly bug reports) and about mining issue repositories.
Recently, several studies have also suggested and partly mined \gls{its} smells. We summarise these areas below.

\subsection{Quality of Issue Reports}
The quality of issue data is a critical prerequisite for effective issue resolution.
Bettenburg et al.~\cite{Bettenburg:FSE:2008} and Zimmerman et al.~\cite{Zimmermann:TSE:2010} first studied gaps between information contained in bug reports and information needed by developers to fix the bugs.
Since then, plenty of follow-up works investigated particular aspects of bug report quality.
Huo et al.~\cite{Huo:ICSME:2014} compared bug reports written by developers vs users and found a significant difference that impacts  prediction models.
Chaparro et al.~\cite{Chaparro:ESECFSE:2019} focused on the quality of steps-to-reproduce in the reports, while Davies and Roper~\cite{Davies:ESEM:2014} focused on observed behaviour and expected results.
These works led to improvements in \gls{its} for creating good bug reports: such as assisting reporters to include relevant and essential information, incentivising good quality bug reports, and merging additional information into existing issue reports~\cite{Breu:CSCW:2010,Just:IEEESymposium:2008}.
Our work is complementary.
Instead of focusing on the reporter perspective, we studied the issue tracking process as it interferes with the work of developers, managers, and product owners.
While some findings from issue quality research are reflected in our investigation (S1.1--S1.3), we studied various issue types beyond bug reports, such as requirements issues and development tasks.

Only a few studies investigated the quality of issues beyond bug reports, notably also \textit{feature requests}.
Heck and Zaidman~\cite{Heck:REJ:2016} defined three levels of feature request completeness: basic, required, and optional.
They found that all feature requests fulfilled the basic completeness, but only 54\% fulfilled ``required''.
Seiler and Paech~\cite{Seiler:REFSQ:2017} ran interviews on the problems with feature requests in \glspl{its} and found that unclear feature descriptions, insufficient traceability, and fragmentation of feature knowledge are common in practice.
Our results confirm their findings and add more context, for example on when details or links to other issues matter in practice and why.
Moreover, we cover other issue types such as User Stories, epics, or tasks---frequently used in modern \glspl{its}~\cite{Montgomery:MSR:2022}.

\subsection{Mining Issue Repositories}
Recent issue datasets suggest that \glspl{its} can include hundreds of thousands of issues with countless comments and links~\cite{Montgomery:MSR:2022, Kallis_2022_NLBSE}.
Several studies have highlighted that large data volumes in \glspl{its} can be overwhelming for manual handling~\cite{Anvik:OOPSLA:2005,Regnell:REFSQ:2008,Fucci:ESEM:2018,Baysal:ICSE:2013}.
Thus, routine tasks such as prioritisation and planning may be become challenging and time-consuming~\cite{Heck:IWPSE:2013,Ernst:EMPIRE:2012, Baysal:ICSE:2013:2}.
Therefore researchers suggested \gls{its} mining approaches to retrieve relevant issues or predict particular fields.
Common goals include issue classification, issue assignment, issue prioritisation, or duplicate detection~\cite{Cavalcanti:JSEP:2014,Zou:TSE:2020}.
Classification aims to identify or correct the issue type (bug, feature request, enhancement etc.) based on issue properties using natural language processing and machine learning (ML)~\cite{Merten:RE:2016,Perez:ICPC:2021}.
Issue assignment aims at predicting an assignee for an issue, for example by analysing the change history, affected components, code ownership~\cite{Xia:WCRE:2013}, issue text and code similarity~\cite{Stanik:ICSME:2018}, or previous contributions~\cite{Rocha:SANER:2016}.
Issue prioritisation aims at predicting the severity, priority, or ranking of issues, based on issue properties~\cite{Menzies:ICSM:2008,Li:ESEM:2022, Tian:WCRE:2012, Lamkanfi:CSMR:2011,Tian:ICSM:2013, Li:ESEM:2022,Izadi:EMSE:2022}.
Our work uses a different research method and focuses on a different perspective.
Instead of data mining, we conducted an interview study with developers and other stakeholders for an-depth understanding of how practitioners use issues and issue properties, as well as problems they encountered.
Specific smells we analysed can be associated to specific mining approaches.

Detecting duplicates as part of issue triage has also been a popular research line.
Researchers evaluated when duplicates are harmful (or not)~\cite{Bettenburg:ICSM:2008,Cavalcanti:CSMR:2010,Cavalcanti:SQJ:2013,Kucuk:SANER:2021,Rakha:EMSE:2016}.
Recently, researchers started looking at other types of issue dependencies too~\cite{Lueders:MSR:2022}, for example to predict parent-children or blocker issues \cite{Lueders:REJ:2023}.
Our work investigates many smells related to issue linking , their occurrence and how problematic they are, thus providing additional details on how to leverage issue dependency mining in practice.

\subsection{Issue Tracking Smells}
This rather new area is the closest to our work, with preliminary studies that presented the starting point for our RQ2. 
Aranda and Venolia~\cite{Aranda:ICSE:2009} conducted a survey and a mining study to extract coordination patterns such as ``forgotten'' and ``close-reopen'' bugs.
Eloranta et al.~\cite{Eloranta:IST:2016} conducted semi-structured interviews to identify 14 agile ``antipatterns''.
Recently, Qamar et al.~\cite{Qamar:SEAA:2021,Qamar:IST:2022} proposed a taxonomy of 12 bug tracking process smells.
They surveyed 30 developers about these smells and mined their occurrences in 8 open-source projects.
Survey respondents also commented on actions they took to avoid the smells.
They observed a considerable amount of the smells in all projects, and the majority of surveyed practitioners agreed with the smells.
Our work covers 19 additional issue smells, including Issue Link and Development Process smells.
We ran hour-long interviews to uncover when and in which context the smells actually may lead to problems and why.
Finally, our goal is to understand the overall challenges in \glspl{its}, including the smells and their management.
Our results discuss possible confounding factors that may explain such correlations.

Tuna et al.~\cite{Tuna:ICSESEIP:2022} studied the 12 smells by Qamar et al.~at JetBrains, surveying 24 developers at this company.
Similarly to our results, they also found the perception of smell severity to vary across smell types, and that smell detection tools are considered useful for only six of the smells.
We studied a larger catalogue of smells by interviewing 26 practitioners in 19 different companies.
Halverson et al.~\cite{Halverson:CSCW:2006} observed collaboration antipatterns in \glspl{its}.
Tamburri et al.~\cite{Tamburri:JISA:2014} defined and evaluated community smells that might lead to unforeseen project cost due to a `suboptimal' community.
They used the term ``community smell'' as organisational and social circumstances which cause mistrust, delays, and uninformed or miscommunicated architectural decision-making.
Palomba et al.~observed that community smells contribute to the intensity of code smells~\cite{Palomba:TSE:2021}. These works motivate our research on issue tracking smells.

In summary, our work investigates a larger number and diversity of smells across multiple aspects of the \gls{its} environment.
We used interviews as research method to allow for a richer qualitative investigation of core problems they were experiencing, compared to conducting a closed-questions survey.
Our participants are from 19 different companies (instead of a case study of one company) and had experience with 18 different \glspl{its} (instead of a focus on one \gls{its}), thus leading to a diverse sample of opinions on problems faced within \glspl{its}.
Our \gls{ta} analysis revealed rich insights, and our closed-coding analysis revealed structured insights into practitioner experiences with our diverse set of \gls{its} smells.

\section{Discussion}\label{sec:disc}

\subsection{Issue Tracker Intelligence Depends}
Our results indicate that there is no ``one size fits all approach'' for intelligent information retrieval and automation in \glspl{its}.
It all depends:
    What information is expected in certain issues and what not;
    What automation is considered to hinder stakeholder's work and what not;
    What assistance is considered helpful;
    How the issue and its attributes should best evolve; Whether issue evolution and associated workflows should be strictly enforced;
    Whether certain discussions and collaboration should take place or not;
    etc.

What may sound like a simple finding is particularly crucial for future research.
For example, automated approaches to retrieve related or duplicate issues, predict the priority, or assign issues, have been so far rather universal.
\textbf{We recommend that researchers} consider these confounding context factors more carefully, as they might strongly impact the evaluation of the proposed intelligent issue tracking solutions.
Our results include context factors such as the \contextName{issue type}, \contextName{stakeholder role}, \contextName{team preference}, and even \contextName{intended outcome}.
How universal the impact of context factors on certain \glspl{its} or types of projects remains unclear.
Future researchers could, for example, use observational studies with developers to control potential biases and clarify how smells should be presented and explained to different stakeholders in the \gls{its}.
For instance, while issue structure visualisation could be helpful for navigating the knowledge graph around an issue, it may seem unnecessary or too complex for people who prefer list visualisation~\cite{Li:REFSQ:2012}.
We hope that our work inspires follow-up studies to formalise, quantify, and measure the impact of the discussed context factors.
The heterogeneity of issue tracking and issue data in practice also sheds light on evaluating ML models trained on issue data.
This heterogeneity influences the issue data, its structure, and semantics.
It seems thus crucial that models to, for example, predict duplicate issues or who should fix a defect should be evaluated across different \glspl{its} and different contexts---not only on popular benchmarking datasets.
We argue that researchers should either precisely scope the \gls{its} context targeted by their ML models or extend the evaluation to heterogeneous issue tracking contexts.

Our results also suggest several \textbf{recommendations for practitioners} concerning 1) the archiving of issue data, 2) the configuration of the trackers, 3) the training of their users, and 4) considerations for multiple stakeholder roles.
\textbf{First}, managers and administrators should carefully consider limiting the size and diversity of issue data, particularly in large long-lasting projects.
Intelligent issue tracking solutions require \textit{meaningful data} as input to their algorithms, and users are only capable of working with so much data in \glspl{its} at once.
Archiving very old issue data could reduce the overload and stress when searching for relevant issue data.
\textbf{Second}, configuring \glspl{its} seems like a highly crucial task that should be revisited regularly, possibly consulting different stakeholders.
If, for instance, old issue types or link types are not used any more, they should be cleaned from the trackers.
This would help clarify situations where the context factor \contextName{issue type} is relevant, or perhaps even the specific task at hand~\cite{Maalej:ASE:2009}. 
Our results suggest that restricting \gls{its} users too much can make them frustrated, but if the degree of freedom is too high with little validation, many mistakes can occur.
Too many issue properties and value options can also become counterproductive.
For example, Herraiz et al. argued for simplifying the issue report form in Eclipse~\cite{Herraiz:MSR:2008}.
The degree of detail and freedom depends on the specific needs and processes of the team or organisation (e.g.,~\contextName{team preference}).
Moreover, most \glspl{its} offer powerful workflow configuration features, for instance automation in Jira.\footnote{\url{https://www.atlassian.com/software/jira/guides/expand-jira/automation-use-cases}}
Our results suggest that some teams seem unaware of (or unwilling to) use these features.
This can become problematic, since a common context factor limiting the existence and impact of smells for our participants was \contextName{\gls{its} configuration}.
\textbf{Third}, training and educating \gls{se} practitioners about issue tracking seems crucial.
Overall, the use of \glspl{its} in the interviewed companies seems ad-hoc (without common principles, rules, or explicit guidelines).
This can be tackled in part by regular training.
For example, training on how to create and evolve issues, what are good and bad practices, and what are powerful options to configure and analyse \glspl{its}.
\textbf{Fourth}, practitioners should be aware of how the differences in roles affects decisions and outcomes with smell detection.
While managers often want more data tracked, more complete data, and more visualisations to traverse and understand the data, developers often want less process, fewer distractions, and easier access to the work they need to conduct.
This balance should not be ignored.

\subsection{Smells Depend on the Context}
Our results highlight that the perceived relevance and severity of 31 smells from the literature are context-dependent.
What is considered risky for some was a  deliberate, intended procedure for others.
For example, the practitioners we interviewed generally agreed that issues should not be delayed for too long, descriptions should be informative, and knowledge should not remain hidden in comments, confirming the relevance and severity of these smells discussed in the literature~\cite{Qamar:SEAA:2021,Qamar:IST:2022,Aranda:ICSE:2009,Prediger_2023_MSc,Lueders_2023_PhDThesis}.
However, the interviewees also noted that certain smells are an \contextName{intended outcome}, such as
    ``often switching properties''~\cite{Halverson:CSCW:2006,Aranda:ICSE:2009,Prediger_2023_MSc},
    ``no comments or too many comments''~\cite{Qamar:SEAA:2021,Qamar:IST:2022,Tamburri:IEEESoftware:2016}, and
    ``issues are assigned to a team''~\cite{Qamar:SEAA:2021,Qamar:IST:2022}.
These results outline the need for context-sensitive smell definitions, analyses, and implementations.

Our \textbf{recommendation for researchers} is to focus on characterising the smells and their severity by analysing the interactions of \gls{its} users and team dynamics over a period of time.
Such learned smells can assist \gls{its} users to externalise and assess their practices and potential smells.
Moreover, it remains unclear whether the \textit{perceptions} of practitioners reflect the \textit{objective} practice.
Future research should investigate the relationship between practitioner perceptions of smells, and the actual measured outcomes of these smells.
Additionally, we believe that certain \gls{its} smells apply across consistent contexts, but more research needs to be conducted to verify this hypothesis.
For example, given the context factors ``company uses acceptance criteria'' and ``acceptance criteria apply to issue types x, y, z'', it is expected that the smell ``no acceptance criteria or too many'' applies within this context, to particular issue types.
Therefore, an additional recommendation for researchers is to study these smells in controlled contexts, and compare the findings to similar but adjusted contexts.
Controlled experiments with \gls{its} smells is likely neither possible nor desired, but multiple case studies investigating similar environments and collecting critical data like issue resolution time and developer opinion on effectiveness can lead to insights on specific contexts.

The \textbf{recommendations for practitioners} include attention to detail when configuring their systems and awareness of context factors when doing so.
First, teams and even single users should be able to configure specific smell detection and management approaches:  not only \textit{what} should be detected as smells but also \textit{at what threshold} and \textit{for whom}.
For example, considering \contextName{team preference} and whether certain workflows have \contextName{supervisor involvement}.
Such configuration can be challenging for administrators, particularly at the setup time of the \gls{its}.
Our work provides guidance for (a) what may matter in which context and (b) how to involve stakeholders in an internal smell configuration study.

The \textbf{recommendations for ITS tool vendors} includes focusing more on contextual factors in the configuration of these environments and automations.
To start, tool vendors should offer a configuration page that allows users to turn certain automations on or off, and configure their sensitivity.
For example, an automation that checks the description length before allowing them to create a new issue, based on conformance to the configuration file.
Next, tool vendors should consider the grouping of users into contexts, such that these groups of users can define shared configurations.
For example, a \textit{team} may consider setting the required Description length to 300 characters.
Finally, tool vendors should configure a way to have overlapping context groups, and transparent layers of overriding configurations.
For example, a company may choose to set the required Description length to 100 characters, just to ensure that no Description is left blank.
However, the team that defined a required Description length of 300 characters should be allowed to override the company configuration because the team is \textit{nested inside} the company context.
Combined, these recommendations for tool vendors gives users greater control over how their \gls{its} reacts to and supports their workflows, across any generic set of nested contexts.
This would directly support the context factors found in this study, including \contextName{workflow design}, \contextName{intended outcome}, and \contextName{\gls{its} configuration}.
All three of these context factors largely mitigate certain smells, so features that support them are likely to improve their underlying systems.

\subsection{Most Problematic Smells}

Based on our findings, the three most problematic smells, each with eleven or more participants stating they are problematic, are as follows.

The smell ``\textit{ignored issue or delayed for too long}'', declared as problematic by 15 participants (and only refuted by one), is the most problematic smell in our dataset.
This smell is problematic because the longer an issue goes unresolved, the harder it is to resolve.
That is because factors like time and people within the organisation affect the resolution of the issue.
After enough time has passed, it might be unclear how to even replicate the reported problem, or who to go to for more information.
If the issue is a requirement, it might not be relevant to the evolved system years or even just months later.
There are many potential reasons for why this smell occurs.
One potential reason is the large number of issues in these \glspl{its}, as reported by our participants.
With so many issues, as well as new ones coming in, issues are bound to be delayed and forgotten.
Another potential reason is when developers are responsible for assigning and working on the issues they want, instead of a centrally managed backlog.
With someone in charge of managing and assigning the backlog, there is then someone responsible for questions like ``are all issues with being worked on or scheduled for the future''.
No one developer is responsible for that, but a manager or product owner can certainly be responsible for that.
One way to address this smell is to automate the reminding of assignees that issues need to be worked on.
For example, an automation could ping an assignee after one week of no activity on an issue, thus reminding them that working on this issue is important.
If the issue is not resolved after three months (for example) the automation could recommend closing or archiving the issue as a way to keep the \gls{its} clean.

The smell ``\textit{properties discussed in the comments but not updated in the issue}'' was declared as problematic by 12 participants and not refuted by any participant.
This smell is particularly problematic because it describes when information is known (and often reported \textit{somewhere}), but is not recorded in the issue.
Therefore, there is a known inconsistency, but it will be unknown to people who only have the information in the issue.
Potential reasons for this include laziness, assumption of responsibility on someone else, and reliance on external tools for discussion.
With so many communication tools in modern workflows, it is easy to see how discussions---and decisions---about issues never propagate back to the issue itself.
In this way, communication has been streamlined within organisations, but at the cost of inconsistent information across \glspl{its} and other tools.
One way to address this smell is to automate the detection of keywords in descriptions and comments while they are being written, and to recommend that users update issue information alongside (or instead of) their textual update.
For example, if a user is writing a comment ``this issue is now resolved, thank you for your help'', but they are not setting the issue's status to ``Resolved'', then the automation could ask them ``do you also want to set the Status to Resolved?'' when they try to save the comment.

The smell ``\textit{no or short (non-informative) description}'' was declared as problematic by 11 participants and refuted by three participants.
This smell is problematic because it might mean that the issue is not actionable, since not enough information is known.
Our participants reported that this happens often, and is the result of laziness or information being stored elsewhere.
Occasionally, the information is contained in a parent issue, or a linked issue, but in either case the issue itself should not still be open, but rather just the issue containing the actual information.
The result of this problematic smell can be a loss of trust in the \gls{its}, since information is often missing.
One way to address this smell is to automate a minimum requirement for Description length, for example, 100 characters or 30 words minimum.
The ``non-informative'' part of this smell is much harder to automatically detect, as ``informative'' is a subjective and context-dependent concept.
However, a naive solution can be implemented using topic modelling and minimum required topics, and more thorough solutions could be implemented using large language models.

\subsection{Key Context Factors}

Certain key context factors prevent or mitigate \gls{its} smells in direct ways, thus potentially providing direct avenues to controlling and minimising smells within an organisation's \gls{its}.
Context factors are normally difficult to influence, since they are embedded in---and even a by-product of---the environment.
For example, \textit{team size} is a context factor that is a consequence of organisational structuring, and not likely something you would change \textit{just} to influence \gls{its} smells.
However, the following key context factors are not traditional context factors, and indeed can be changed to influence the quality outcomes of \glspl{its}, thus offering potential as a useful tool for organisations looking to address \gls{its} smells.

The first key context factor is \contextName{workflow design}.
Our study found that many smells were caught early or prevented entirely as part of an organisation's standard workflow, such as regular meetings.
The most relevant smells for this key context factor are
    S1.8 ``\textit{non-assignee resolved issue}'',
    S3.2 ``\textit{too many complex issues are assigned to the same sprint}'', and
    S1.9 ``\textit{Ignored issue or delayed for too long}''.
S1.9 is also the most problematic smell, thus putting additional emphasis on the need to prevent or mitigate it.
Organisations should seek to avoid it (as it is the most problematic), and one way to do that is through the key context factor \contextName{workflow design}.
Our \textbf{recommendation for practitioners} is to implement one of the following workflows, all of which were described by our participants as mitigating this smell.
The first workflow is to have management (manager, product owner, etc.) assign issues to developers, and not allow self-assignment.
While this puts responsibility on management and takes away some developer agency, it grants the opportunity for management to maintain a constant overview of all issues, including those that are not receiving attention, and thus allows for adjustment and prioritisation above what any one developer would feel responsible for.
The next workflow is to have regular team meetings to discuss open issues (whether assigned or not), thus creating a team awareness of all open issues, and supporting an open dialogue and prioritisation of the entire backlog.
The final recommended workflow is to give someone the responsibility of maintaining and curating the backlog of open issues, thus encouraging a relevant, trimmed, and curated backlog for developers to choose from.
If open issues are left too long, unassigned, then the curator can increase the priority/severity until a developer chooses to work on the issue.

The second key context factor is \contextName{intended outcome}.
Our study found that many smells are, in fact, not a smell, but rather an intended or expected outcome of a certain workflow.
The most relevant smells for this key context factor are
    S1.10 ``\textit{no comments or too many comments}'',
    S1.4 ``\textit{issues are assigned to a team}'',
    S1.5 ``\textit{often switching properties}'',
    S3.3 ``\textit{issues are missing an estimate}'', and
    S1.3 ``\textit{issues are missing properties}''.
On the one hand, this is a commentary about research more than it is practice.
Our \textbf{recommendation for researchers} is to further investigate these smells and their classification as smells based on evidence from industry.
We need a growing body of evidence that describes which smells are occurring and problematic in industry, to enable informed decisions about the importance of smells.
On the other hand, this is also a commentary on how practitioners approach the quality of their \glspl{its}.
Our \textbf{recommendation for practitioners} is to view quality of their \glspl{its} as a \textit{selective process}, based on how their organisation functions.
Not all smells will be relevant or important to every organisation, and so decisions need to be made to focus on what matters.
As a start, the evidence from our study has shown that the five smells listed above (S1.10, S1.4, S1.5, S3.3, and S1.3) are not relevant to many of the studied organisations.
Thus, perhaps they are less likely to be relevant to other organisations.
For the remainder of the smells, it would be beneficial to do a detailed check of which smells seem important to their organisation, before making any changes or deploying any automations to support and report the identification of these smells within their organisation.
Our \textbf{recommendation for tool vendors} is to offer configuration options that allow organisations to mark certain smells as ``intended outcome'' so they can ignore them, and focus on the smells that do matter to them.
We describe this in more detail in the next Discussion point.

The third and final key context factor is \contextName{\gls{its} configuration}.
Our study found that many smells are prevented entirely by the implementation of certain \gls{its} features.
The most relevant smells for this key context factor are
    S1.4 ``\textit{issues are assigned to a team}'',
    S1.7 ``\textit{no link to commit}'',
    S1.3 ``\textit{issues are missing properties}'', and
    S2.1 ``\textit{issue without any links}''.
\glspl{its} are complex systems which often allow for many configurations and automations, many of which can support the prevention of certain \gls{its} smells.
Configuration, for example, can force the filling in of desired fields, thus preventing S1.3 entirely.
Automations also support processes such as automatic linking of issues to their resolving commit, thus preventing S1.7.
Our participants said that prevention is better than correction because it is less work for them.
Our \textbf{recommendation for practitioners} is to carefully configure their \gls{its} to enforce behaviour they think is essential, and automate procedures that are common and simple, such as linking Bug Reports to resolving GIT commits.
Our \textbf{recommendation for tool vendors} is to integrate better onboarding processes to support the creation and maintenance of configurations and automations.
Our participants also mentioned that configuring these \glspl{its} is difficult, thus leading to a sense of helplessness when using these systems, despite having administrative control.

\section{Threats to Validity}\label{sec:threats}
As with every empirical study, there are limitations which might have influenced our results.

\textbf{Internal Validity.} 
Our main research method was semi-structured interviews.
Particularly as we did not record the sessions (for the criticality of the discussion and for better spontaneous engagement), observer bias represents a risk for our work.
We might have misunderstood the interviewees or missed important points in our reporting.
To mitigate this risk, each session was conducted by two interviewers.
Both of them took notes at all times and asked for clarifications whenever needed.
We also discussed each interview session afterwards to mitigate the threat of memory recall.
Another potential threat is the observer-expectancy effect, where the interviewers elicit the statements they are expecting or looking for.
We took multiple measures to mitigate this risk, as discussed in Section \ref{sec:method}.
Notably, we explicitly encouraged participants to disagree and share their personal opinions.
We also had two interviewers in each session, posed our questions as neutral and general as possible, and exposed participants to information only when necessary (e.g.~showing them the smells).
Due to participants' disagreements as well as the heterogeneous perceptions gathered, we think this bias remains minor.
Similarly, the results are based on the subjective statements of practitioners.
In theory, what people do might diverge from what they say, introducing a self-reporting bias.
Therefore, triangulation with observation studies or artefact analysis will likely lead to stronger evidence and more insights.
Finally, to address potential researcher bias, the \gls{ta} was conducted by three separate authors, is fully documented, and each sentence traced to the findings.

\textbf{External Validity.}
Compared to previous studies, our smell list is extensive, covering issue properties, linking, and process smells.
However, as our smell collection process was neither complete nor systematic, we certainly have not received feedback on all types of \gls{its} smells.
Our goal was to discuss various \gls{its} smells with practitioners rather than compiling a full smell catalogue.
Additional smells might lead to different perceptions and results.
Our sample includes ``only'' 26 experienced software practitioners.
We think that this is a rather large diverse sample comparable to seminal qualitative studies in \gls{se}.
The sample provided enough diversity and redundancy to derive our findings.
However, we refrain from claiming generalisability of the results to software practitioners, and this never was our goal.
Nevertheless, the diversity of the sample, involved 19 companies, and redundancy of observation give us reason to believe that the overall trends hold true.
To achieve quantifiable and generalisable results, follow-up studies (such as surveys or experiments with practitioners) are needed.
Our qualitative findings serve as starting hypotheses and variables to measure for such studies.

\section{Conclusion}\label{sec:conc}
We identified common challenges faced by \gls{its} users by interviewing 26 experienced practitioners from diverse domains.
The identified challenges include the efficient retrieval of relevant information from the \gls{its}, maintaining and navigating workflows within the \gls{its}, and managing the \gls{its} usage on an organisational level.
Default search features and universal issue property recommenders only address these challenges in part.
Our work motivates researching ``intelligent'' \gls{its} approaches based on mining issue data and considering context factors such as issue types, practitioners roles, previous interaction with the issues, and diversity of the \gls{its}.
Our results also highlight how software teams are striking the balance between automation and flexibility in \glspl{its}.
Our study explains how and when the informal and flexible nature of \glspl{its} can lead to many smells related to issue properties, issue links, and development processes.
Our work explores whether and how these smells interfere with the work of developers, managers, and product owners.
Generally, delayed or ignored issues, missing commits in issue reports, and missing information are considered problematic.
However, other smells, such as often switching properties, are sometimes followed deliberately with certain workflow logic or flexibility in mind.
Whether the smells are problematic or not seems to depend on particular context factors, suggesting that tools for detecting and managing issue smells should be customisable to the teams and individual stakeholders.
Whether this customisation needs to be done manually or can be automated is a further open question for future work.
Future research should investigate and possibly quantify context causality, that is which context factors affect which smells, and in what way, thus creating more specific and actionable recommendations for practitioners.
Practitioners should identify which \gls{its} smells would be beneficial for their contexts, and implement a mix of mitigation techniques including configuration, automation, and workflow adjustments.
Finally, \gls{its} tool vendors should increase the configurability of \glspl{its} and expand automation options, while also simplifying the onboarding and usage of these more advanced features---towards a better ITS experience for developers and other stakeholders.

\begin{acks}
We thank all study participants for their time and valuable input.
We would like also to thank Nina Prediger for her support in conducting some of the interviews in the first round.
\end{acks}

\bibliographystyle{ACM-Reference-Format}
\bibliography{main}

\end{document}